\documentclass[12pt,a4paper]{article}  
\usepackage{cite}
\usepackage{epsfig}
\usepackage{graphicx}
\usepackage{amsmath}
\usepackage{amssymb}
\usepackage{color}  
\usepackage[T1]{fontenc}
\usepackage[english]{babel}

\usepackage{a41}
\usepackage{color}
\usepackage[rflt]{floatflt}
\usepackage{float}
\usepackage{slashed}

\DeclareMathOperator{\arctanh}{arctanh}

\usepackage{array}

\usepackage[english]{babel}

\usepackage{url}

 \newcommand{\GeV}{\mathrm{GeV}}

 \newcommand{\MS}{\overline{\sf MS}}

\newcommand{\NN}{\nonumber}

\newcommand{\Li}{{\rm Li}}
\newcommand{\HA}{{\rm H}}

\newcommand{\Mvec}{{\rm\bf M}}

\newcommand{\ep}{\varepsilon}

\newcommand{\dd}{\mathrm{d}}

\usepackage{rotating}

\usepackage{graphicx}

\newcounter{mmacnt}
\def\restartmma{\setcounter{mmacnt}{0}}
\restartmma \catcode`|=\active
\def|#1|{\mathrm{#1}}
\catcode`|=12
\newenvironment{mma}{
 \par\smallskip
 \catcode`|=\active
 \parskip=0pt\parindent=0pt 
 \small
 \def\In##1\\{%
\def\linebreak{\hfill\break\null\qquad}%
\refstepcounter{mmacnt}
\hangindent=2.5em\hangafter=0
\leavevmode
\llap{\tiny\sffamily n[\arabic{mmacnt}]:=\kern.5em}%
\mathversion{bold}\footnotesize$\displaystyle##1$\normalsize
\mathversion{normal}\par
 }%
 \def\Print##1\\{%
\def\linebreak{\hfill\break}%
\hangindent=2.5em\hangafter=0
\leavevmode ##1\par}%
 \def\Out##1\\{%
\def\linebreak{$\hfill\break\null\hfill$}%
\kern\abovedisplayskip\par
\hangindent=2.5em\hangafter=0
\leavevmode
\llap{\tiny\sffamily Out[\arabic{mmacnt}]=\kern.5em}
\footnotesize$\displaystyle##1$\normalsize\hfill\null\par
\kern\belowdisplayskip
 }%
 \def\Warning##1##2\\{%
\def\linebreak{\hfill\break}%
\hangindent=2.5em\hangafter=0
\leavevmode
{\scriptsize##1 : ##2}\par}%
}{%
 \par\smallskip
}

\newenvironment{fshaded}{%
\MakeFramed {\FrameRestore}
}%
{\endMakeFramed}

\allowdisplaybreaks[4]

\begin{document}
\setlength{\baselineskip}{0.515cm}
\sloppy
\thispagestyle{empty}
\begin{flushleft}

DESY 20--051 
\\
DO--TH 20/03 \\
TTP 20--013\\
SAGEX--20--07--E
\\
April 2020\\
\end{flushleft}

\mbox{}
\vspace*{\fill}
\begin{center}

{\LARGE\bf The Two-mass Contribution to}

\vspace*{7mm}
{\LARGE\bf \boldmath the Three-Loop Polarized Gluonic}

\vspace*{2mm}
{\LARGE\bf \boldmath Operator Matrix Element $A_{gg,Q}^{(3)}$}

\vspace{3cm}
\large
J.~Ablinger$^a$, 
J.~Bl\"umlein$^b$, 
A.~De Freitas$^b$,
A.~Goedicke$^c$, 

\vspace*{2mm}
M.~Saragnese$^b$, 
C.~Schneider$^a$,  and  K.~Sch\"onwald$^{b,c}$ 

\vspace{1.cm}
\normalsize
{\it $^a$~
Johannes Kepler University Linz,\\
Research Institute for Symbolic Computation (RISC),\\
Altenberger Stra\ss{}e 69, A--4040, Linz, Austria}

\vspace*{3mm}
{\it  $^b$ Deutsches Elektronen--Synchrotron, DESY,}\\
{\it  Platanenallee 6, D-15738 Zeuthen, Germany}

\vspace*{3mm}
{\it  $^c$ Institut f\"ur Theoretische Teilchenphysik
Campus S\"ud,} \\ 
{\it Karlsruher Institut f\"ur Technologie (KIT) D-76128 Karlsruhe, Germany}
\\

\end{center}
\normalsize
\vspace{\fill}
\begin{abstract}
\noindent
We compute the two-mass contributions to the polarized massive operator matrix element $A_{gg,Q}^{(3)}$ 
at third order in the strong coupling constant $\alpha_s$ in Quantum Chromodynamics analytically. These corrections 
are important ingredients for the matching relations in the variable flavor number scheme and for the calculation 
of Wilson coefficients in deep--inelastic scattering in the asymptotic regime $Q^2 \gg m_c^2, m_b^2$. The analytic 
result is expressed in terms of nested harmonic, generalized harmonic, cyclotomic and binomial sums in $N$-space 
and by iterated integrals involving square-root valued arguments in $z$ space, as functions of the mass ratio. 
Numerical results are presented. New two--scale iterative integrals are calculated.
\end{abstract}

\vspace*{\fill}
\noindent
\numberwithin{equation}{section}
\newpage 
\section{Introduction}
\label{sec:1}

\vspace*{1mm}
\noindent
The massive operator matrix elements (OMEs) of twist--2 operators emerge in the asymptotic representation of the
massive Wilson coefficients in deeply inelastic scattering \cite{Buza:1995ie} in the region $Q^2 \gg m_c^2, m_b^2$. 
Here $m_i = m_{c(b)}$ are the charm and bottom quark masses\footnote{We will use the on--shell scheme for the mass 
renormalization. The transformation to the $\overline{\sf MS}$ scheme is straightforward, cf.~\cite{Ablinger:2014vwa}.} and 
$Q^2$ denotes the virtuality of the deeply inelastic 
process. They also appear  as the matching coefficients in the variable flavor number scheme (VFNS). Various of 
these OMEs have been calculated already in the unpolarized and polarized case in
Refs.~\cite{Buza:1995ie,
Buza:1996xr,
Buza:1996wv,
Bierenbaum:2007qe,
Bierenbaum:2007pn,
Bierenbaum:2008yu,
Ablinger:2010ty,
Ablinger:2014vwa,
Ablinger:2014nga,
Ablinger:2014uka,
Behring:2014eya,
Blumlein:2014fqa,
Ablinger:2014lka,
Behring:2015zaa,
Behring:2016hpa,
Ablinger:2017ptf,
Ablinger:2016kgz,
Blumlein:2016xcy,
Ablinger:2019gpu,
Ablinger:2019etw,
Blumlein:2019zux,
Blumlein:2019qze}.
Furthermore, two--mass corrections contribute from the two--loop corrections onward, \cite{Blumlein:2018jfm}, with 
one--particle irreducible contributions starting at the three--loop level \cite{Ablinger:2011pb,Ablinger:2012qj,
Ablinger:2018brx,Ablinger:2017err,Ablinger:2019gpu,Ablinger:2017xml}.
At 3--loop order also the contributing anomalous dimensions are obtained as a by-product of the calculation of the 
massive OMEs \cite{Ablinger:2014vwa,Ablinger:2014nga,Ablinger:2017tan,Behring:2019tus}, 
confirming the results obtained in massless calculations \cite{Moch:2004pa,Vogt:2004mw,Moch:2014sna}.

In performing the different calculation steps we use the algorithms encoded in the packages \texttt{Sigma} \cite{SIG1, SIG2},
{\tt HarmonicSums} \cite{Vermaseren:1998uu,Blumlein:1998if,Ablinger:2014rba,Ablinger:2010kw, Ablinger:2013hcp, 
Ablinger:2011te,Ablinger:2013cf,Ablinger:2014bra}, {\tt EvaluateMultiSums} and {\tt 
SumProduction}\cite{EMSSP}.
The computation follows the methods used in the unpolarized case \cite{Ablinger:2018brx}. 

In the present paper we calculate the polarized three--loop massive operator matrix element 
$A_{gg,Q}^{(3)}$ using dimensional regularization in $d = 4 + \ep$ dimensions. 
The computation is performed in the Larin scheme 
\cite{Larin:1993tq}.\footnote{For the 
anomalous dimensions in the Larin scheme see \cite{Moch:2014sna,Behring:2019tus}.} 
The renormalization procedure is the 
same as in the unpolarized case, cf.~\cite{Ablinger:2017err,Ablinger:2018brx}.
We perform the calculation of the massive OME first in Mellin $N$ space and switch then to momentum 
fraction $z$ space by performing an inverse Mellin transform analytically. Various calculation 
techniques applied are described in Ref.~\cite{Blumlein:2018cms}.
In both spaces the results are expressed by special functions introduced in Refs. 
\cite{Vermaseren:1998uu,Blumlein:1998if,Moch:2001zr,Ablinger:2011te,Ablinger:2013cf,Ablinger:2014bra,Remiddi:1999ew}.
Starting from the result in $N$ space, the inverse Mellin transform to $z$ space is performed by finding 
a 
recurrence relation satisfied by the various sums, and by using the properties of the inverse Mellin transform 
to build a differential equation that the Mellin inverse must satisfy, see Refs.~\cite{Ablinger:2016lzr,Ablinger:2018cja,
Ablinger:2018pwq}. The differential equation is then solved. These methods are implemented in 
\texttt{HarmonicSums}.

The paper is organized as follows. In Section~\ref{sec:2} we summarize the details of the calculation. In many technical
aspects we will refer to the calculation in the unpolarized case \cite{Ablinger:2018brx} and to Ref.~\cite{Ablinger:2017err}.
In Section~\ref{sec:3}, fixed moments are calculated for the constant part of the unrenormalized massive OME by a 
different method to have an \'etalon to check the present results in Mellin $N$ space. The result in Mellin $N$ space is 
derived in Section~\ref{sec:4}. In Section~\ref{sec:5} we present the transformation to $z$ space, and in 
Section~\ref{sec:6} we derive numerical results comparing the size of the two--mass effects to the complete 
contributions of $O(C_F T_F^2)$. Section~\ref{sec:7} contains the conclusions. In Appendix~\ref{sec:A}, a series of new 
iterative integrals are presented and relations between different iterated integrals are listed in 
Appendix~\ref{sec:B} beyond those given in Ref.~\cite{Ablinger:2018brx} before. 
\section{Details of the Calculation}
\label{sec:2}

\begin{figure}[ht]
\begin{center}
\begin{minipage}[c]{0.20\linewidth}
  \includegraphics[width=1\textwidth]{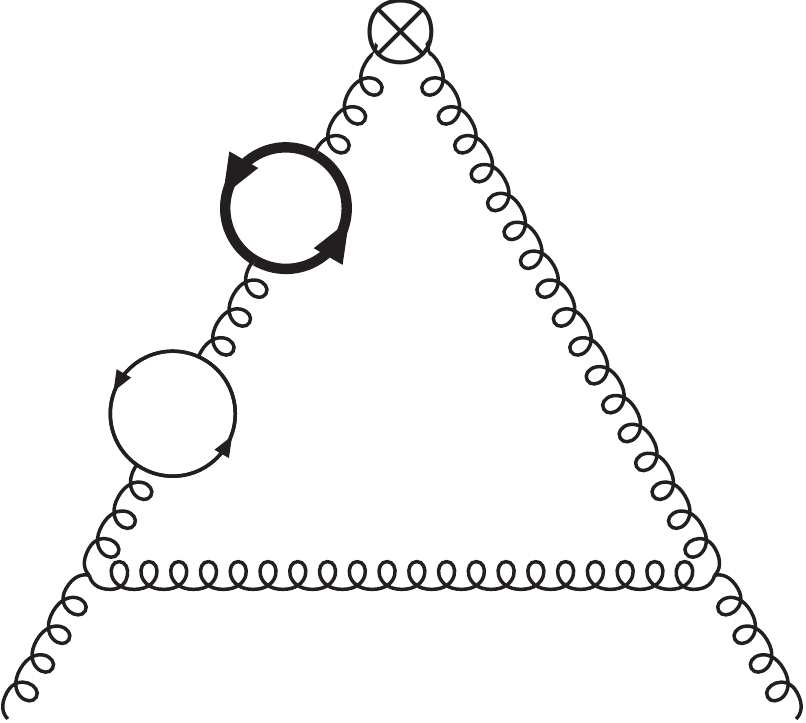}
\vspace*{-11mm}
\begin{center}
{\footnotesize (1)}
\end{center}
\end{minipage}
\hspace*{1mm}
\begin{minipage}[c]{0.20\linewidth}
  \includegraphics[width=1\textwidth]{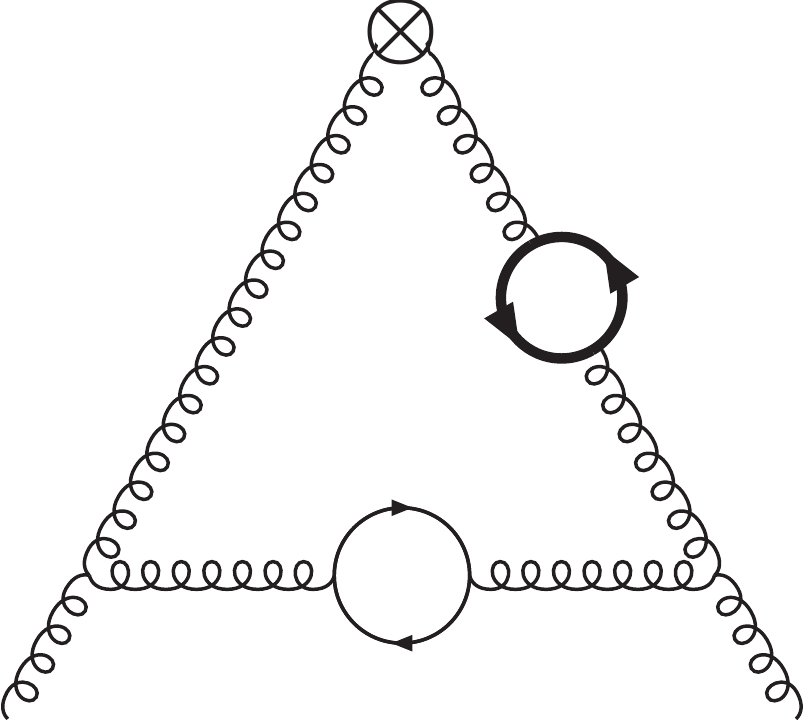}
\vspace*{-11mm}
\begin{center}
{\footnotesize (2)}
\end{center}
\end{minipage}
\hspace*{1mm}
\begin{minipage}[c]{0.20\linewidth}
  \includegraphics[width=1\textwidth]{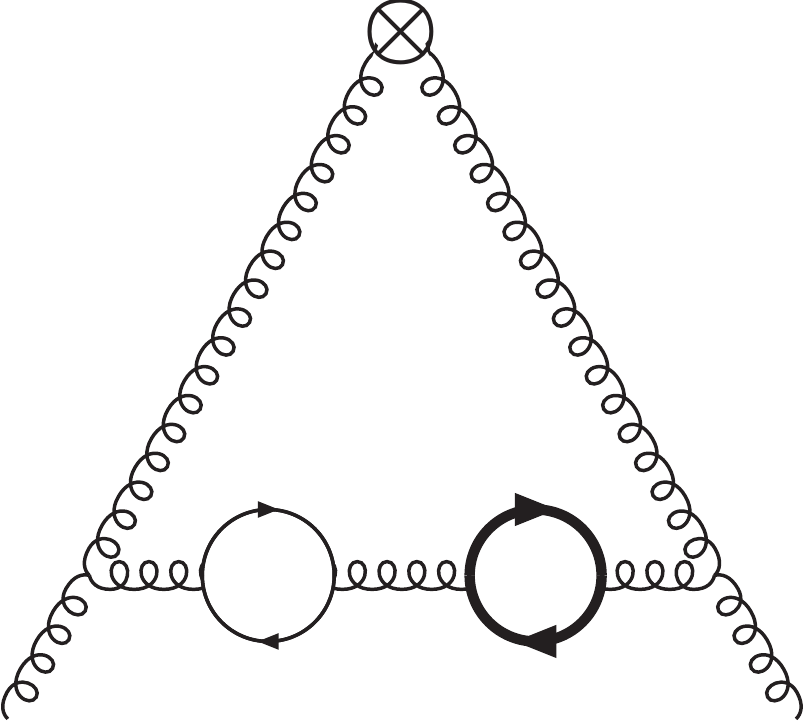}
\vspace*{-11mm}
\begin{center}
{\footnotesize (3)}
\end{center}
\end{minipage}
\hspace*{1mm}
\begin{minipage}[c]{0.20\linewidth}
  \includegraphics[width=1\textwidth]{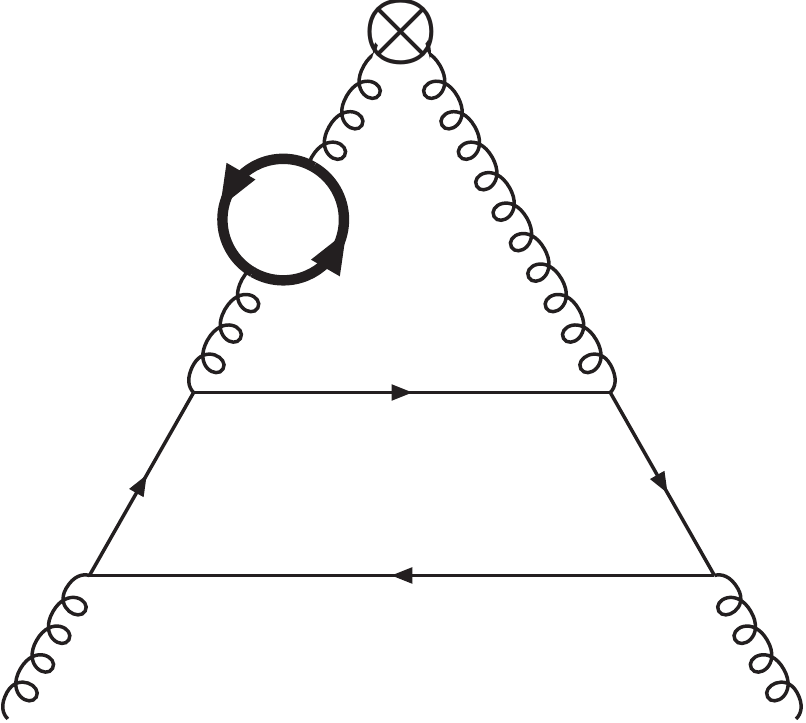}
\vspace*{-11mm}
\begin{center}
{\footnotesize (4)}
\end{center}
\end{minipage}

\vspace*{5mm}

\begin{minipage}[c]{0.20\linewidth}
  \includegraphics[width=1\textwidth]{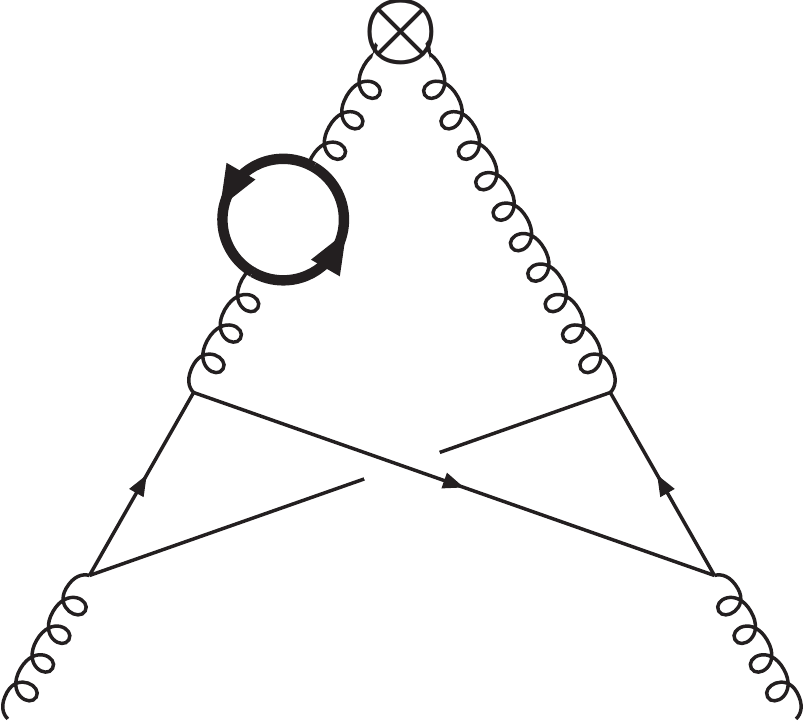}
\vspace*{-11mm}
\begin{center}
{\footnotesize (5)}
\end{center}
\end{minipage}
\hspace*{1mm}
\begin{minipage}[c]{0.20\linewidth}
  \includegraphics[width=1\textwidth]{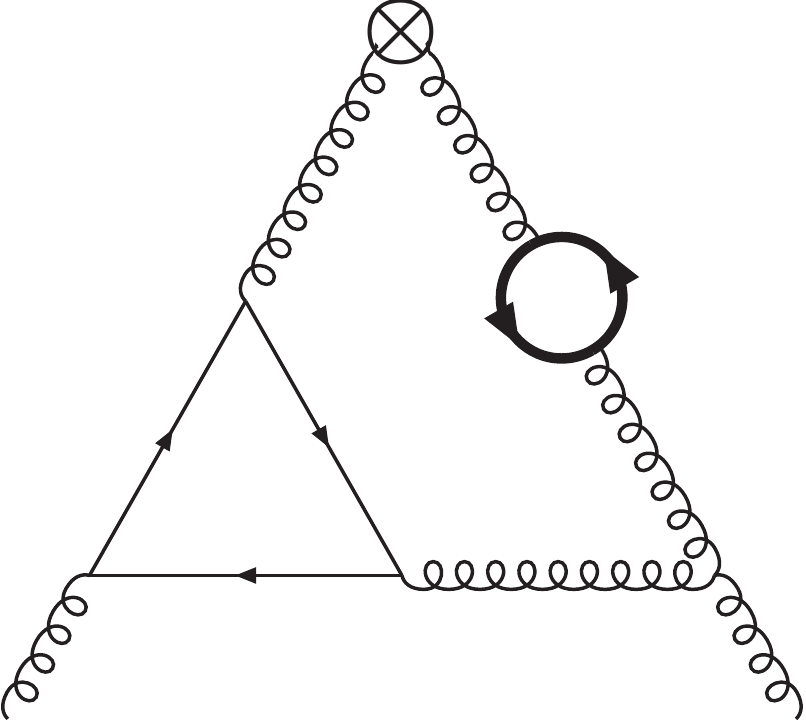}
\vspace*{-11mm}
\begin{center}
{\footnotesize (6)}
\end{center}
\end{minipage}
\hspace*{1mm}
\begin{minipage}[c]{0.20\linewidth}
  \includegraphics[width=1\textwidth]{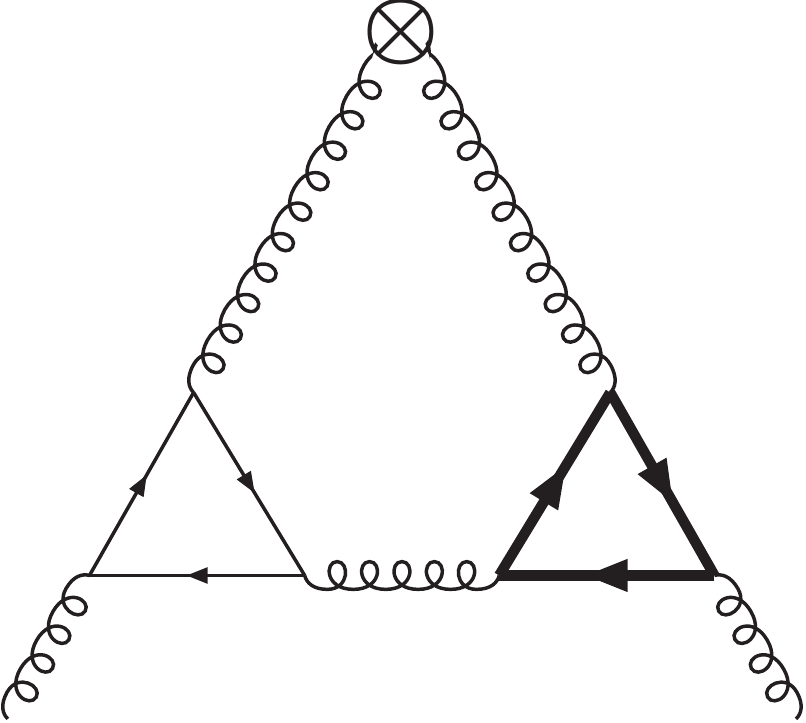}
\vspace*{-11mm}
\begin{center}
{\footnotesize (7)}
\end{center}
\end{minipage}
\hspace*{1mm}
\begin{minipage}[c]{0.20\linewidth}
  \includegraphics[width=1\textwidth]{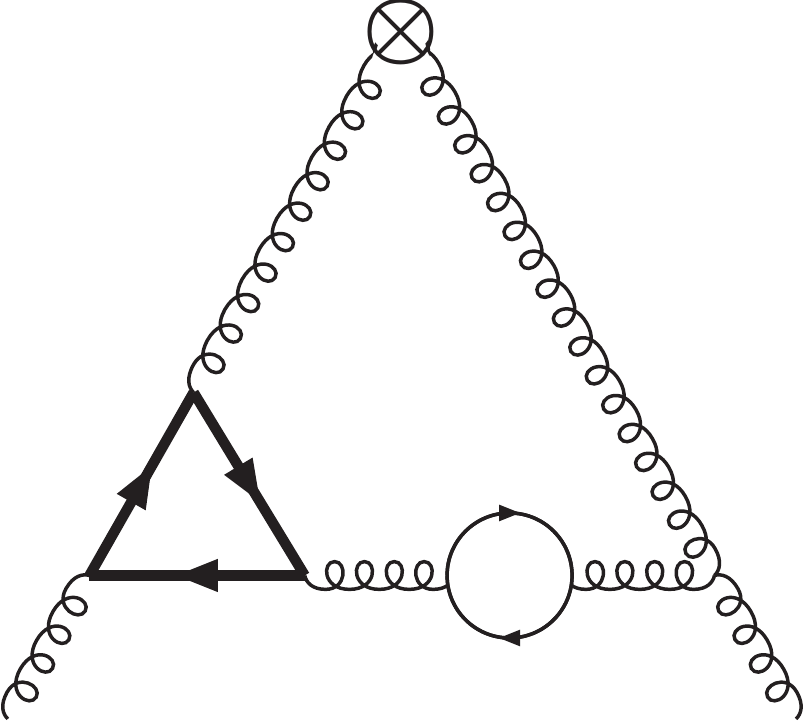}
\vspace*{-11mm}
\begin{center}
{\footnotesize (8)}
\end{center}
\end{minipage}

\vspace*{5mm}

\begin{minipage}[c]{0.20\linewidth}
  \includegraphics[width=1\textwidth]{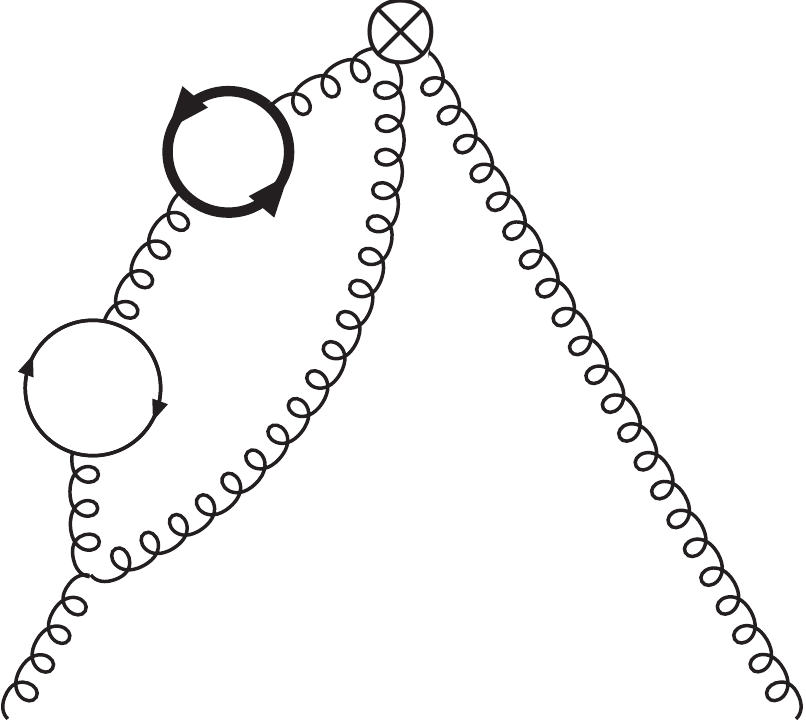}
\vspace*{-11mm}
\begin{center}
{\footnotesize (9)}
\end{center}
\end{minipage}
\hspace*{1mm}
\begin{minipage}[c]{0.20\linewidth}
  \includegraphics[width=1\textwidth]{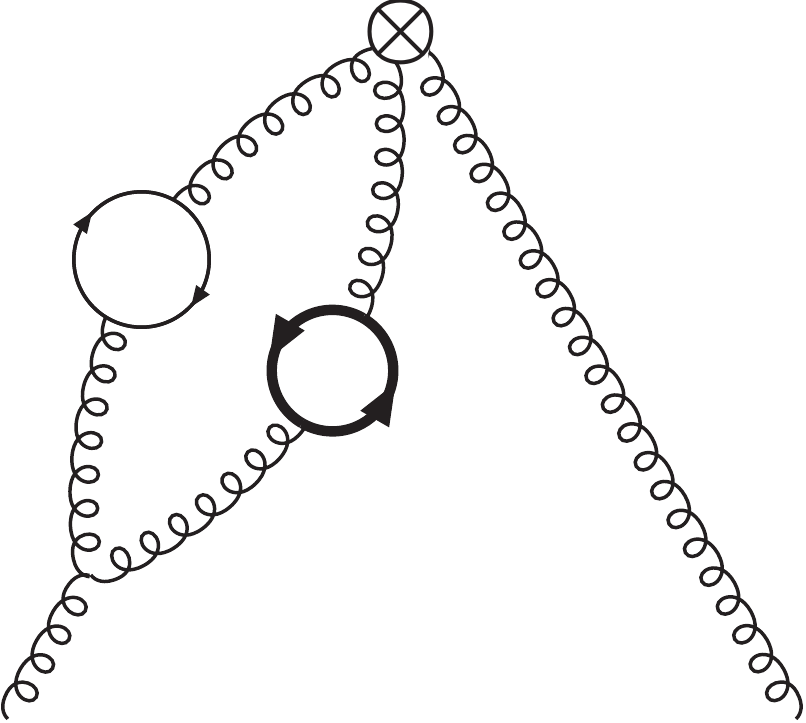}
\vspace*{-11mm}
\begin{center}
{\footnotesize (10)}
\end{center}
\end{minipage}
\hspace*{1mm}
\begin{minipage}[c]{0.20\linewidth}
  \includegraphics[width=1\textwidth]{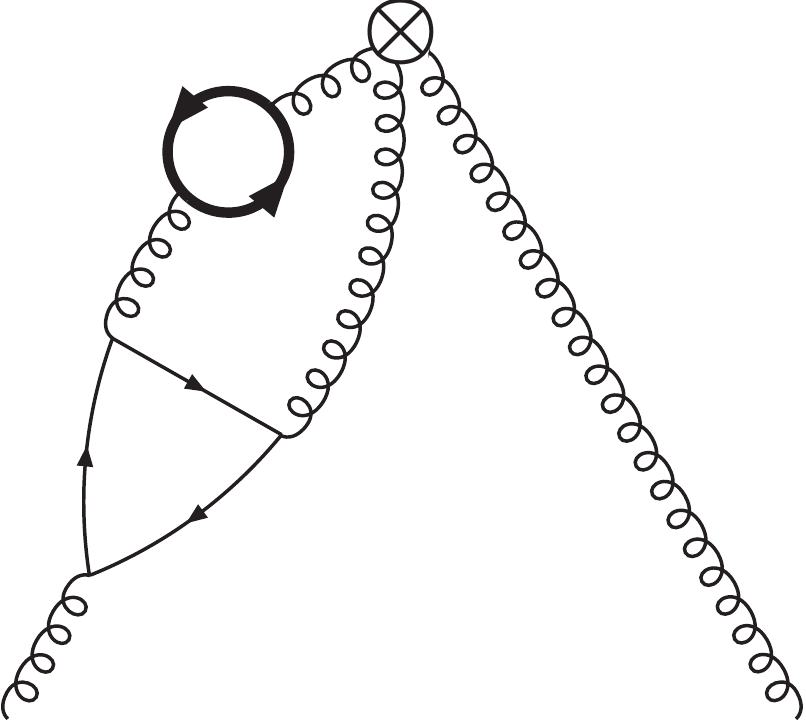}
\vspace*{-11mm}
\begin{center}
{\footnotesize (11)}
\end{center}
\end{minipage}
\hspace*{1mm}
\caption{\small \sf The 11 different topologies for $\tilde{A}_{gg,Q}^{(3)}$. Curly lines: gluons; thin arrow 
lines: lighter massive quark; thick arrow lines: heavier massive quark; the symbol $\otimes$ represents the corresponding 
local operator insertion, cf.~\cite{Mertig:1995ny} for the related Feynman rules.
\label{fig:1}}
\end{center}
\end{figure}

\vspace*{1mm}
\noindent
As it is well--known from Ref.~\cite{Ablinger:2018brx}, the new contribution is the constant part of the
unrenormalized polarized two mass OME $\hat{A}_{gg,Q}^{(3)}$, since the pole contributions are predicted 
by the renormalization group equations for the given quantity and are thus determined by known lower order terms, which 
are already known. We first compute the irreducible contributions 
to the unrenormalized polarized OME $\hat{A}_{gg,Q}^{(3)}$. The reducible contributions were given in 
Ref.~\cite{Blumlein:2017mtk}. There are seventy different diagrams contributing. Since many of them have 
the same value, one may group them into eleven equivalence classes, see Table~\ref{TAB1}. The number is 
smaller than in the unpolarized case, because some of the diagrams being present there vanish in the 
polarized case. One example graph for each class is depicted in Figure~\ref{fig:1}.

We use the notation
\begin{equation}
\eta=\frac{m_{2}^{2}}{m_{1}^{2}}<1,\qquad L_{1}=\log\left(\frac{m_{1}^{2}}{\mu^{2}}\right),\qquad 
L_{2}=\log\left(\frac{m_{2}^{2}}{\mu^{2}}\right),
\end{equation}
where $m_i$ are the unrenormalized quark masses, and $\mu$ the renormalization and factorization scale.
The total result for $\hat{A}_{gg,Q}^{(3)}$ is symmetric interchanging the masses $m_1 \leftrightarrow 
m_2$.
For diagrams which are not symmetric, it is always possible to recover one mass assignment from the 
other using the simultaneous exchange
\begin{equation}\label{eq:symmM1M2}
m_{1}\leftrightarrow m_{2},\qquad\eta\rightarrow\frac{1}{\eta}.
\end{equation}
However, as a cross--check on the computation, for all asymmetric diagrams, except diagram 8 and 11, we choose 
to compute the two mass assignments independently, using a different Mellin--Barnes decomposition for 
the two cases, and we explicitly check that the results are related by Eq.~\eqref{eq:symmM1M2}.
\begin{table}[h!]
\begin{center}
\begin{tabular}{|cc|cc|}
\hline 
Diagram & Multiplicity & Diagram & Multiplicity\tabularnewline
\hline 
\hline 
1 & 3 & 7 & 4\\
\hline 
2 & 2 & 8 & 4\\
\hline 
3 & 1 & 9 & 2\\
\hline 
4 & 4 & 10 & 1\\
\hline 
5 & 2 & 11 & 4\\
\hline 
6 & 8 &  \multicolumn{1}{c}{} & \multicolumn{1}{c}{}\\
\cline{1-2} \cline{2-2}
\end{tabular}
\caption{\sf \small Symmetry properties under $m_{1}\leftrightarrow m_{2}$ and multiplicity of the 
eleven diagrams of 
Figure~\ref{fig:1}. The multiplicity refers to the sum of the two mass assignments.}
\label{TAB1}
\end{center}
\end{table}

The Feynman diagrams are generated using {\tt QGRAF} \cite{QGRAF} and the Dirac algebra is performed by using 
{\tt FORM} \cite{FORM}. The Feynman diagrams contain local operator insertions, cf.~\cite{Ablinger:2017err}, and 
the corresponding Feynman rules were given in \cite{Mertig:1995ny,Yndurain:1999ui}; see also \cite{Behring:2019tus}.
We work in the Larin scheme~\cite{Larin:1993tq}. The unrenormalized OME is obtained by applying to the 
Green's 
function for the eleven irreducible diagrams the projector~\cite{Bierenbaum:2009mv,Klein:2009ig}
\begin{equation}\label{eq:projector}
A_{gg,Q}=\frac{\delta^{ab}}{N_c^2-1}
                    \frac{1}{(d-2)(d-3)}
                    (\Delta\cdot p)^{-N-1}\ep^{\mu\nu\rho\sigma}
                    \Delta \hat{G}^{ab}_{Q,\mu\nu}
                    \Delta_{\rho}p_{\sigma}\,,
\end{equation}
where $p$ is the external momentum, with $p^2 = 0$,  $\Delta$ is 
a lightlike momentum, $N_c$ is the number of colors, and $\varepsilon^{\alpha_1 \alpha_2 \alpha_3 \alpha_4}$ denotes the Levi--Civita 
tensor. The one--particle 
reducible terms have still to be added, cf.~\cite{Ablinger:2017err}. Ghost diagrams do not contribute to the amplitudes.

The renormalized expression for the OME is given by \cite{Ablinger:2018brx}
\begin{eqnarray}
\label{eq:renOME3}
\tilde{A}_{gg,Q}^{(3), \MS}
&=&
\Bigg\{
\frac{25}{24} \beta_{0,Q}^2 \gamma_{gg}^{(0)}
+\frac{25}{12} \beta_{0} \beta_{0,Q}^2
+\frac{9}{2} \beta_{0,Q}^3
+\frac{23}{96} \hat{\gamma}_{qg}^{(0)} \beta_{0,Q} \gamma_{gq}^{(0)}
\Biggr\} \left(L_1^3+L_2^3\right)
+\Biggl\{
\frac{1}{8} \hat{\gamma}_{qg}^{(0)} \beta_{0,Q} \gamma_{gq}^{(0)}
\NN\\&&
+\beta_{0,Q}^2 \gamma_{gg}^{(0)}
+2 \beta_{0} \beta_{0,Q}^2
+6 \beta_{0,Q}^3
\Biggr\} \left(L_1^2 L_2+L_2^2 L_1\right)
+\Biggl\{
-\frac{1}{4} \beta_{1,Q} \beta_{0,Q}
+\frac{13}{16} \beta_{0,Q} \hat{\gamma}_{gg}^{(1)}
\NN\\&&
+\frac{29}{4} \delta m_1^{(-1)} \beta_{0,Q}^2
-\frac{1}{64} \hat{\gamma}_{qg}^{(0)} \hat{\gamma}_{gq}^{(1)}
\Biggr\} \left(L_1^2+L_2^2\right)
+8 L_2 L_1 \delta m_1^{(-1)} \beta_{0,Q}^2
+\Biggl\{
 \frac{9}{4} \beta_{0} \beta_{0,Q}^2 \zeta_2
\NN\\&&
+\frac{27}{2} \beta_{0,Q}^3 \zeta_2
-3 \beta_{0,Q} \tilde{\delta}m_2^{(-1)}
+\frac{9}{8} \zeta_2 \beta_{0,Q}^2 \gamma_{gg}^{(0)}
+12 \delta m_1^{(0)} \beta_{0,Q}^2
+\frac{3}{32} \beta_{0,Q} \zeta_2 \gamma_{gq}^{(0)} \hat{\gamma}_{qg}^{(0)}
\NN\\&&
+6 \beta_{0,Q} a_{gg,Q}^{(2)}
\Biggr\} \left(L_2+L_1\right)
-\frac{1}{32} \hat{\gamma}_{qg}^{(0)} \zeta_2 \hat{\gamma}_{gq}^{(1)}
+\frac{1}{8} \beta_{0,Q} \zeta_2 \hat{\gamma}_{gg}^{(1)}
+\frac{1}{3} \beta_{0} \beta_{0,Q}^2 \zeta_3
+12 \beta_{0,Q} \overline{a}_{gg,Q}^{(2)}
\NN\\&&
+6 \beta_{0,Q}^3 \zeta_3+16 \delta m_1^{(1)} \beta_{0,Q}^2
+\frac{1}{6} \beta_{0,Q}^2 \zeta_3 \gamma_{gg}^{(0)}
-2 \beta_{0,Q} \left(
\tilde{\delta} {m_2}^{1,(0)}
+\tilde{\delta} {m_2}^{2,(0)}
\right)
\NN\\&&
+\frac{9}{2} \delta m_1^{(-1)} \beta_{0,Q}^2 \zeta_2
-\frac{1}{24} \zeta_3 \beta_{0,Q} \gamma_{gq}^{(0)} \hat{\gamma}_{qg}^{(0)}
-\frac{1}{2} \zeta_2 \beta_{0,Q} \beta_{1,Q}
+\tilde{a}_{gg,Q}^{(3)}\left(m_1^2,m_2^2,\mu^2\right)~. \label{Agg3QMSren}
\end{eqnarray}
Here, $\beta_i$ and $\beta_{Q,i}$ are expansion coefficients of the $\overline{\sf MS}$ scheme and 
background field
$\beta$ functions, $\gamma_{ij}^{(k)}$ are anomalous dimensions, $\delta m_k^{(l)}$ are expansion parameters of the
unrenormalized quark masses and $a_{ij}^{(k)}$ and $\overline{a}_{ij}^{(k)}$ are lower loop expansion coefficients 
of massive OMEs. $\zeta_k,~k \geq 2$ denotes the Riemann $\zeta$ function at integer arguments \cite{Blumlein:2009cf}, see 
Refs.~\cite{Ablinger:2017err,Ablinger:2018brx} for details.

Because the number of classes of diagrams is small, we choose to derive one Feynman parameterization per diagram and we 
do not cancel numerator and denominator structures, nor do we reduce to master integrals. This allows us to deal with 
a uniform denominator structure which we find to be simpler to treat. The additional complexity 
potentially induced 
on the numerators by this choice still turns out to be manageable. For quark bubble insertions, we use the relation
\begin{eqnarray}
    \Pi^{\mu\nu}_{ab} ( k ) = -i \frac{ 8 T_F g^2}{(4\pi)^{d/2}} \delta_{ab} ( k^2 g^{\mu\nu} - k^\mu k^\nu ) 
\int\limits_{0}^{1} \text{d}x \frac{ \Gamma(2-d/2) \left( x(1-x) \right)^{d/2-1} }
{ \left( -k^2 + \frac{m^2}{x(1-x)} \right)^{2-d/2} } .
\end{eqnarray}
Once the Feynman parameterization has been obtained for the whole diagram, the integrals over the loop momenta 
are performed one after another, using the tensor identities
\begin{eqnarray}
 \int \frac{\dd^d k}{(2\pi)^d}  k_{\mu_1} k_{\mu_2} f(k^2) 
&=&  \frac{g_{\mu_1\mu_2}}{d} \int \frac{\dd^d k}{(2\pi)^d}  k^2 f(k^2), \\
\int \frac{\dd^d k}{(2\pi)^d}  k_{\mu_1} k_{\mu_2} k_{\mu_3} k_{\mu_4} f(k^2)  
&=& \frac{S_{\mu_1\mu_2\mu_3\mu_4}}{d(d+2)}  \int \frac{\dd^d k}{(2\pi)^d} (k^2)^2 f(k^2),
\nonumber\\
\\
\int \frac{\dd^d k}{(2\pi)^d}  k_{\mu_1} k_{\mu_2} k_{\mu_3} k_{\mu_4} k_{\mu_5} k_{\mu_6} f(k^2)  
&=& \frac{S_{\mu_1\mu_2\mu_3\mu_4\mu_5\mu_6}}{d(d+2)(d+4)} 
\int \frac{\dd^d k}{(2\pi)^d}  (k^2)^3 f(k^2),
\end{eqnarray}
with the symmetric tensors
\begin{eqnarray}
S_{\mu_1\mu_2\mu_3\mu_4} &=& g_{\mu_1\mu_2}g_{\mu_3\mu_4} + g_{\mu_1\mu_3}g_{\mu_2\mu_4} 
+ g_{\mu_1\mu_4} g_{\mu_2\mu_3} 
\\
S_{\mu_1\mu_2\mu_3\mu_4\mu_5\mu_6} &=& 
g_{\mu_1\mu_2} 
\left[
  g_{\mu_3\mu_4} g_{\mu_5\mu_6} 
+ g_{\mu_3\mu_5} g_{\mu_4\mu_6}
+ g_{\mu_3\mu_6} g_{\mu_4\mu_5}\right]
\nonumber\\
&&
+g_{\mu_1\mu_3} 
\left[
  g_{\mu_2\mu_4} g_{\mu_5\mu_6} 
+ g_{\mu_2\mu_5} g_{\mu_4\mu_6}
+ g_{\mu_2\mu_6} g_{\mu_4\mu_5}\right]
\nonumber\\
&&
+g_{\mu_1\mu_4} 
\left[
  g_{\mu_2\mu_3} g_{\mu_5\mu_6} 
+ g_{\mu_2\mu_5} g_{\mu_3\mu_6}
+ g_{\mu_2\mu_6} g_{\mu_3\mu_5}\right]
\nonumber\\
&&
+g_{\mu_1\mu_5} 
\left[
  g_{\mu_2\mu_3} g_{\mu_4\mu_6} 
+ g_{\mu_2\mu_4} g_{\mu_3\mu_6}
+ g_{\mu_2\mu_6} g_{\mu_3\mu_4}\right]
\nonumber\\
&&
+g_{\mu_1\mu_6} 
\left[
  g_{\mu_2\mu_3} g_{\mu_4\mu_5} 
+ g_{\mu_2\mu_4} g_{\mu_3\mu_5}
+ g_{\mu_2\mu_5} g_{\mu_3\mu_4}\right].
\end{eqnarray}
The scalar integrals can then be performed using the relation
\begin{eqnarray}
\int \frac{\dd^d k}{(2\pi)^d} \frac{ (k^2)^m }{ ( k^2 + R^2 )^n } 
&=& \frac{1}{(4\pi)^{d/2}} \frac{\Gamma(n-m-d/2)}{\Gamma(n)} \frac{\Gamma(m+d/2)}{\Gamma(d/2)} 
\left( R^2 \right)^{m-n+d/2} .
\end{eqnarray}
After this step, only integrations over the Feynman parameters remain. They can always be cast into the 
form
\begin{equation}\label{eq:feynIntegrals}
    \prod\limits_{i=1}^{j} \int\limits_0^1 \mathrm{d} x_i \ x_i^{a_i} (1-x_i)^{b_i} \ R_0^N \ \left[ R_1 
\ m_1^2 + R_2 \ m_2^2 \right]^{-s},
\end{equation} 
where $R_0$ is a polynomial in the Feynman parameters $x_i$, and $R_1$ and $R_2$ are rational functions 
in $x_i$.
At this point the polynomial $R_0$ can be treated by applying the binomial theorem (multiple times if necessary)
\begin{equation}
    (A + B)^N = \sum\limits_{i=0}^{N} \binom{N}{i} A^i B^{N-i}
\label{eq:I2}
\end{equation}
and a Mellin--Barnes decomposition 
\cite{MB1a,MB1b,MB2,MB3,MB4} is applied to 
the factor $\left[ R_1 \ m_1^2 + R_2 \ m_2^2 \right]^{-s}$, 
\begin{equation}
\frac{1}{(A+B)^{s}} = \frac{1}{2\pi i} \frac{1}{\Gamma (s)} B^{-s} \int\limits_{-i\infty}^{+i\infty} 
\mathrm{d} \sigma \left( \frac{A}{B} \right)^{\sigma} \Gamma (-\sigma) \Gamma( \sigma + s ) ,
\end{equation}
where the integration contour must separate the ascending from the descending poles in the $\Gamma$--functions. After 
appropriately closing the contour at infinity, this integral is turned into a number of infinite sums using the 
residue theorem. At this point the integrals over $x_i$ can always be expressed in terms of Beta functions.

The effect of these steps is to transform expressions having the form~\eqref{eq:feynIntegrals}, i.e. integrals over 
Feynman parameters, into nested sums. This sum representation can be handled by \texttt{Sigma}, 
\texttt{EvaluateMultiSums}, \texttt{SumProduction}, assisted by  \texttt{HarmonicSums}.

We checked our computation using the packages \texttt{MB}~\cite{MB} and \texttt{MBResolve}~\cite{MBr} 
to resolve the singularity structure of the Mellin--Barnes integrals numerically and to compute the 
$\varepsilon$-poles of the 
various diagrams 
for fixed values of $N$, and compared them to the general $N$ results obtained using \texttt{Sigma}.

The target function space for the result is that of harmonic sums~\cite{Vermaseren:1998uu,Blumlein:1998if}
\begin{eqnarray}\label{Equ:HarmonicSumsIntro}
S_{b,\vec{a}}(N) &=& \sum_{k=1}^N \frac{({\rm sign}(b))^k}{k^{|b|}} S_{\vec{a}}(k),
a_i,b \in \mathbb{Z} \backslash \{0\}, N \in \mathbb{N}, S_\emptyset 
=1,
\end{eqnarray}
generalized harmonic sums~\cite{Moch:2001zr,Ablinger:2013cf,Ablinger:2013jta}
\begin{eqnarray}
S_{b,\vec{a}}(d,\vec{c})(N) &=& \sum_{k=1}^N \frac{d^k}{k^{|b|}} S_{\vec{a}}(\vec{c})(k),
a_i,b \in \mathbb{N} \backslash \{0\}, c_i,d \in \mathbb{C}(\sqrt{\eta}) \backslash \{0\}, N \in \mathbb{N}, 
S_\emptyset 
=1,
\end{eqnarray}
cyclotomic sums~\cite{Ablinger:2011te} and binomial sums~\cite{Ablinger:2014bra}. Here the index $d$ 
denotes a function of 
$\sqrt{\eta}$. Harmonic polylogarithms~\cite{Remiddi:1999ew} will also appear in the result. The procedure adopted here 
is the same as used in Ref.~\cite{Ablinger:2018brx}, where the corresponding unpolarized OME was 
computed. There, the reader can find further details and also explicit examples about the computational 
method.

The renormalization procedure is the same as in the unpolarized case, including the removal of collinear singularities,
cf.~\cite{Ablinger:2017err}. 

All structures of the massive OME are predicted by the renormalization or are available from lower loop results,
except for the constant term in the two--mass case, $\tilde{a}_{gg,Q}^{(3)}$, of the unrenormalized three--loop 
OME.

First we turn to the calculation of fixed moments of this quantity.
\section{Fixed moments of \boldmath $\tilde{a}_{gg,Q}^{(3)}$}
\label{sec:3}
Here we present the results for the fixed moments $N=3$ and 5 of the massive OME including the reducible contributions. 
They were computed using the packages \texttt{Q2E/EXP}~\cite{Harlander:1997zb,Seidensticker:1999bb} up to 
$\mathcal{O}(\eta^2)$ and $\mathcal{O}(\eta)$, respectively. In what follows we use the abbreviation
\begin{equation}
L_\eta=\ln(\eta).
\end{equation}
The moments are given by
\begin{eqnarray}
\tilde{a}_{gg,Q}^{(3)}(N=3) &=& \textcolor{blue}{C_F T_F^2} \Bigg\{
-\frac{444341}{11664}
+\bigg(
        -\frac{1031}{18}
        +\frac{25 L_1}{9}
        +\frac{25 L_2}{9}
\bigg) \zeta_2
-\frac{220}{81} \zeta_3
-\frac{22121}{648} L_1
\nonumber\\&&
-\frac{5635}{108} L_1^2
+\frac{155}{81} L_1^3
-\frac{76411}{1944} L_2
-\frac{1822}{27} L_1 L_2
+\frac{10}{27} L_1^2 L_2
-\frac{5635}{108} L_2^2
\nonumber\\&&
+\frac{50}{27} L_1 L_2^2
+\frac{115}{81} L_2^3
+\eta\bigg(
\frac{5528}{675}+\frac{112}{45}L_{\eta } \bigg)
+\eta^2\bigg(
-\frac{6202718}{1157625}+\frac{32804 }{11025}L_{\eta }
\nonumber\\&&
-\frac{116 }{105}L_{\eta }^2 \bigg)
\bigg\}
+\textcolor{blue}{C_A T_F^2}\bigg\{
-\frac{2146957}{8748}
+\bigg(
        -\frac{3070}{81}
        -\frac{560}{9} L_1
        -\frac{560}{9} L_2
\bigg) \zeta_2
\nonumber\\&&
+\frac{320}{81} \zeta_3
-\frac{94403}{486} L_1
-\frac{3005}{81} L_1^2
-\frac{2320}{81} L_1^3
-\frac{16781}{162} L_2
-\frac{3200}{81} L_1 L_2
\nonumber\\&&
-\frac{800}{27} L_1^2 L_2
-\frac{3005}{81} L_2^2
-\frac{1120}{27} L_1 L_2^2
-\frac{2000}{81} L_2^3
+\eta\bigg(
-\frac{1869176}{30375}
\nonumber\\&&
+\frac{61346}{2025} L_{\eta }
-\frac{313}{135} L_{\eta }^2
\bigg)
+\eta^2\bigg(
-\frac{94073888}{10418625}
+\frac{486314}{99225} L_{\eta }
-\frac{1126}{945} L_{\eta }^2
\bigg)
\bigg\}
\nonumber\\&&
+\textcolor{blue}{T_F^3}\bigg\{
\bigg(
        32 L_1   
        +32 L_2
\bigg) \zeta_2
+\frac{128}{9} \zeta_3
+\frac{32}{3} L_1^3
+\frac{64}{3} L_1^2 L_2
+\frac{64}{3} L_1 L_2^2
+\frac{32}{3} L_2^3
\bigg\}
\nonumber\\&&
+\mathcal{O}\Big(\eta^3 L_\eta^3\Big)~,\\
\tilde{a}_{gg,Q}^{(3)}(N=5) &=& \textcolor{blue}{C_F T_F^2} \Bigg\{
-\frac{20217949}{455625}
+\bigg(
        -\frac{12976}{225}
        +\frac{56}{45} L_1
        +\frac{56}{45} L_2
\bigg) \zeta_2
-\frac{2464}{2025} \zeta_3
\nonumber\\&&
-\frac{1202098}{30375} L_1
-\frac{177256}{3375} L_1^2
+\frac{1736}{2025} L_1^3
-\frac{411926}{10125} L_2
-\frac{229408}{3375} L_1 L_2
\nonumber\\&&
+\frac{112}{675} L_1^2 L_2
-\frac{177256}{3375} L_2^2
+\frac{112}{135} L_1 L_2^2
+\frac{1288}{2025} L_2^3
+\eta\bigg(
\frac{52768}{10125}+\frac{1984}{675} L_{\eta }
\bigg)
\bigg\}
\nonumber\\&&
+\textcolor{blue}{C_A T_F^2}\bigg\{
-\frac{466789273}{1366875}
+\bigg(
        -\frac{125176}{2025}
        -\frac{3724}{45} L_1
        -\frac{3724}{45} L_2
\bigg) \zeta_2
+\frac{2128}{405} \zeta_3
\nonumber\\&&
-\frac{2733284}{10125} L_1
-\frac{120092}{2025} L_1^2
-\frac{15428}{405} L_1^3
-\frac{4547582}{30375} L_2
-\frac{135344}{2025} L_1 L_2
\nonumber\\&&
-\frac{1064}{27} L_1^2 L_2
-\frac{120092}{2025} L_2^2
-\frac{7448}{135} L_1 L_2^2
-\frac{2660}{81} L_2^3
+\eta\bigg(
-\frac{21339914}{253125}
\nonumber\\&&
+\frac{73016}{1875} L_{\eta }
-\frac{11146}{3375} L_{\eta }^2
\bigg)
\bigg\}
+\textcolor{blue}{T_F^3}\bigg\{
\big(
        32 L_1
        +32 L_2
\big) \zeta_2
+\frac{128}{9} \zeta_3
+\frac{32}{3} L_1^3
\nonumber\\&&
+\frac{64}{3} L_1^2 L_2
+\frac{64}{3} L_1 L_2^2
+\frac{32}{3} L_2^3
\bigg\}
+\mathcal{O}\Big(\eta^2 L_\eta^2\Big)~,
\end{eqnarray}
where $C_F = (N_c^2-1)/(2 N_c), T_F = 1/2, C_A = N_c$ for $SU(N_c)$ and $N_c  = 3$ in the case of QCD.
The calculational methods used in the packages {\tt Q2E/EXP} are very different of those of the
present calculation. Therefore, these results, in form of an expansion in $\eta$, can be used
for comparison. 

We have compared also the fixed moments of each diagram to their calculation at general values of $N$.
For diagrams 1--8 we also calculated the first moment.
\section{The Result in Mellin Space}
\label{sec:4}

\vspace*{1mm}
\noindent
In Mellin $N$ space 
the renormalized massive OME $\tilde{A}_{gg,Q}^{(3), \overline{\rm MS}}$ is given 
by
\begin{eqnarray} 
\tilde{A}_{gg,Q}^{(3), \overline{\rm MS}} &=& \biggl[-\frac{184}{9} 
\textcolor{blue}{C_F T_F^2} h_1
        +\frac{400}{27} \textcolor{blue}{C_A T_F^2} h_2
        -\frac{32}{3} \textcolor{blue}{T_F^3}
\biggr]
\left(
        L_1^3
        +L_2^3
\right)
\nonumber \\ &&
+\biggl[-\frac{32}{3} \textcolor{blue}{C_F T_F^2} h_1
        +\frac{128}{9} \textcolor{blue}{C_A T_F^2} h_2
        -\frac{128}{9} \textcolor{blue}{T_F^3}
\biggr]
\left(
        L_1^2 L_2
        +L_1 L_2^2
\right)
\nonumber \\ &&
+\frac{256}{3} \textcolor{blue}{C_F T_F^2} L_1 L_2
+\biggl[\textcolor{blue}{C_A T_F^2} \biggl(
                -\frac{4 T_2}{27 N^2 (N+1)^2}
                +\frac{1040}{27} S_1
        \biggr)
\nonumber \\ &&
        +\textcolor{blue}{C_F T_F^2} \biggl(
                \frac{8 T_4}{9 N^3 (N+1)^3}
                -\frac{16}{3} h_1 S_1
        \biggr)
\biggr]
\left(
        L_1^2
        +L_2^2
\right)
\nonumber \\ &&
+ \biggl[
        \textcolor{blue}{C_A T_F^2} \biggl(
                -\frac{16 T_5}{27 N^3 (N+1)^3}
                +\frac{32 (56 N+47)}{27 (N+1)} S_1
                +\frac{112}{3} \zeta_2 h_2
        \biggr)
\nonumber \\ &&
        + \textcolor{blue}{C_F T_F^2} \biggl(
                -\frac{32 T_6}{3 N^4 (N+1)^4}
                -40 \zeta_2 h_1
        \biggr)
        -32 T_F^3 \zeta_2
\biggr]
\left(L_1+L_2\right)
\nonumber \\ && 
+ \textcolor{blue}{C_F T_F^2} \biggl\{
        \frac{4 T_8}{9 N^5 (N+1)^5 \eta }
        +\frac{16 T_9 \zeta_2}{9 N^3 (N+1)^3 \eta ^2}
\nonumber \\ && 
        -\frac{64}{3}
\big(\sqrt{\eta }+1\big)^2 \big(
                \eta -\sqrt{\eta }+1\big) 
\biggl[\frac{1}{\eta^2} \HA_{-1,0}\biggl(\frac{1}{\sqrt{\eta }}\biggr)
+\HA_{-1,0}\big(\sqrt{\eta }\big)
\biggr]
\nonumber \\ && 
        + \frac{64}{3}
\big(\sqrt{\eta }-1\big)^2 \big(\eta +\sqrt{\eta }+1\big) 
\biggl[\frac{1}{\eta ^2} \HA_{1,0}\biggl(\frac{1}{\sqrt{\eta }}\biggr)
+\HA_{1,0}\big(\sqrt{\eta }\big)
\biggr]
\nonumber \\ &&
        -\frac{32}{3} h_1 \biggl(
                \zeta_2 S_1
               +\frac{5}{3} \zeta_3
        \biggr)
        +\frac{32 (\eta^2 -1)}{3 \eta } \ln(\eta )
        +\frac{16 \big(
                \eta ^4+1\big)}{3 \eta ^2} \ln^2(\eta )
\biggr\}
\nonumber \\ &&
+\textcolor{blue}{C_A T_F^2} \biggl\{
        -\frac{16 T_7}{81 N^4 (N+1)^4}
        +\biggl[
                \frac{32 T_1}{81 (N+1)^2}
                +\frac{1120}{27} \zeta_2
        \biggr] S_1
\nonumber \\ &&
        +\frac{16 S_1^2}{3 (N+1)}
        -\frac{16 (2 N+1)}{3 (N+1)} S_2
        -\frac{8 T_3 \zeta_2}{27 N^2 (N+1)^2}
        +\frac{448}{27} \zeta_3 h_2
\biggr\}
\nonumber \\ && 
-\frac{128}{9} \textcolor{blue}{T_F^3} \zeta_3 + \tilde{a}_{gg,Q}^{(3)}\left(m_1^2,m_2^2,\mu^2\right) \, 
,
\end{eqnarray}
where $h_1$ and $h_2$ are given by
\begin{eqnarray}
h_1 &=& \frac{(N-1) (N+2)}{N^2 (N+1)^2}, \\
h_2 &=& S_1-\frac{2}{N (N+1)},
\end{eqnarray}
and the polynomials $T_i$ read
\begin{eqnarray}
T_1 &=& 328 N^2+584 N+283, \\
T_2 &=& 171 N^4+342 N^3+847 N^2+676 N-156, \\
T_3 &=& 99 N^4+198 N^3+463 N^2+364 N-84, \\
T_4 &=& 66 N^6+198 N^5+169 N^4+80 N^3+140 N^2-47 N-78, \\
T_5 &=& 15 N^6+45 N^5+374 N^4+601 N^3+161 N^2-24 N+36, \\
T_6 &=& 2 N^8+8 N^7+18 N^6+20 N^5+2 N^4-3 N^2-9 N-6, \\
T_7 &=& 3 N^8+12 N^7+2080 N^6+5568 N^5+4602 N^4+1138 N^3 \nonumber \\ &&
-3 N^2-36 N-108, \\
T_8 &=& -144 \eta +72 \eta ^2 N^{10}+247 \eta  N^{10}+72 N^{10}+360 \eta ^2 N^9+1235 \eta  N^9 \nonumber \\ &&
+360 N^9+720 \eta ^2 N^8+2182 \eta  N^8+720 N^8+720 \eta ^2 N^7+1606 \eta  N^7 \nonumber \\ &&
+720 N^7+360 \eta ^2 N^6+875 \eta  N^6+360 N^6+72 \eta ^2 N^5+1183 \eta  N^5 \nonumber \\ &&
+72 N^5+1152 \eta  N^4+216 \eta  N^3 -288 \eta  N^2-360 \eta  N, \\
T_9 &=& -42 \eta ^2-36 \eta ^{7/2} N^6-36 \eta ^{5/2} N^6-36 \eta ^{3/2} N^6+12 \eta ^4 N^6+57 \eta ^2 N^6 \nonumber \\ &&
-36 \sqrt{\eta } N^6+12 N^6-108 \eta ^{7/2} N^5-108 \eta ^{5/2} N^5-108 \eta ^{3/2} N^5 \nonumber \\ &&
+36 \eta ^4 N^5+171 \eta ^2 N^5-108 \sqrt{\eta } N^5+36 N^5-108 \eta ^{7/2} N^4 \nonumber \\ &&
-108 \eta ^{5/2} N^4-108 \eta ^{3/2} N^4+36 \eta ^4 N^4+160 \eta ^2 N^4-108 \sqrt{\eta } N^4 \nonumber \\ &&
+36 N^4-36 \eta ^{7/2} N^3-36 \eta ^{5/2} N^3-36 \eta ^{3/2} N^3+12 \eta ^4 N^3+71 \eta ^2 N^3 \nonumber \\ &&
-36 \sqrt{\eta } N^3+12 N^3+68 \eta ^2 N^2-29 \eta ^2 N.
\end{eqnarray}
Here, the functions $\HA_{\vec{a}}$ denote the harmonic polylogarithms \cite{Remiddi:1999ew}
\begin{eqnarray}
\HA_{b,\vec{a}}(x) = \int_0^x dz f_b(z) \HA_{\vec{a}}(z),~~f_b(z) \in \left\{f_0 = \frac{1}{z}, 
f_1 = \frac{1}{1-z}, 
f_{-1} = \frac{1}{1+z}\right\},~~H_\emptyset = 1. 
\end{eqnarray}
Later also cyclotomic harmonic polylogarithms \cite{Ablinger:2011te} will contribute. To account for 
their letters, a two--index structure is needed, e.g.
\begin{eqnarray}
f_{4,0} = \frac{1}{1+x^2},~~~~~f_{4,1} = \frac{x}{1+x^2}.
\end{eqnarray}
The denominators are cyclotomic polynomials with $p_3 = 1+x+x^2, p_4 = 1+ x^2, p_5 = 1+x+x^2+x^3+x^4, p_6 = 1-x+x^2$ etc.
The maximal degree of the numerator for the $k$th cyclotomic denominator polynomial is given by $l = \varphi(k)$, where $\varphi$ denotes
Euler's totient function.

All contributions except of the constant part of the unrenormalized OME, 
$\tilde{a}_{gg,Q}^{(3)}\left(m_1^2,m_2^2,\mu^2\right)$, are predicted by the renormalization procedure,
which provides a check of the present calculation.

The Mellin--Barnes and binomial representations of the individual Feynman diagrams imply sum representations
which can be summed using the technologies provided by the packages {\tt Sigma, EvaluateMultisums} and {\tt SumProduction}.
Here we distinguish the cases of the real parameter $\eta$ and $1/\eta$. The computation time for the calculation
amounted to several months. The solution is given in terms of harmonic sums, generalized harmonic sums and finite binomial 
and inverse binomial sums \cite{Ablinger:2014bra} over generalized sums. 

In the following, whenever the argument of a harmonic polylogarithm is omitted, it is implied to be $\eta$, and whenever the 
argument of a harmonic sum is omitted, it is implied to be $N$.

One obtains 


where the polynomials $P_i$ are
\begin{eqnarray}
P_1&=&-27
-36 N^2
-36 N (-2+\eta )
+54 \eta 
+5 \eta ^2
\\
P_2&=&5
-2 \big(
        -11-18 N+18 N^2\big) \eta 
+5 \eta ^2
\\
P_3&=&372 \eta 
+N^2 \big(
        5-102 \eta +5 \eta ^2\big)
+N \big(
        5-66 \eta +5 \eta ^2\big)
\\
P_4&=&-5
+18 (-3+2 N) \eta 
+9 (-3+2 N) (-1+2 N) \eta ^2
\\
P_5&=&-80 \big(
        5+22 \eta +5 \eta ^2\big)
+3 N^2 \big(
        71-46 \eta +71 \eta ^2\big)
+3 N \big(
        167+18 \eta +167 \eta ^2\big)
\\
P_6&=&-80 \big(
        5+22 \eta +5 \eta ^2\big)
+3 N^2 \big(
        111+64 \eta +111 \eta ^2\big)
+3 N \big(
        207+128 \eta +207 \eta ^2\big)
\\
P_7&=&-(2+N) \Big[
        27
        +36 N^2
        +36 N (-2+\eta )
        -54 \eta 
        -5 \eta ^2
\Big]\\
P_8&=&24
-48 N
-18 N^3
+5 N^2 (9+\eta )
\\
P_9&=&(2+N) (-4+3 N) (-2+3 N) (1+\eta )\\
P_{10}&=&-24 \eta 
+48 N \eta 
+18 N^3 \eta 
-5 N^2 (1+9 \eta )
\\
P_{11}&=&9
+160 \eta 
+50 N^2 \eta 
+140 N^3 \eta 
+9 \eta ^2
-9 N \big(
        7+10 \eta +7 \eta ^2\big)
\\
P_{12}&=&(2+N) \Big[
        11
        -54 \eta 
        +11 \eta ^2
        +18 N^2 \big(
                1+\eta ^2\big)
        -36 N \big(
                1-\eta +\eta ^2\big)
\Big]\\
P_{13}&=&(2+N) \Big[
        -5
        +18 (-3+2 N) \eta 
        +9 (-3+2 N) (-1+2 N) \eta ^2
\Big]\\
P_{14}&=&168-536 N-407 N^2-278 N^3-167 N^4\\
P_{15}&=&-(2+N) \Big[
        -24
        +48 N
        +18 N^3
        -5 N^2 (9+\eta )
\Big]\\
P_{16}&=&-96+370 N+277 N^2+12 N^3+9 N^4\\
P_{17}&=&48-27 N+263 N^2+634 N^3+344 N^4\\
P_{18}&=&-1632+7072 N+8611 N^2+3078 N^3+1539 N^4\\
P_{19}&=&-20 N^4 (10+81 \eta )
-4 N^3 \big(
        41+855 \eta +135 \eta ^2\big)
+N \big(
        41+855 \eta -765 \eta ^2-87 \eta ^3\big)
\nonumber\\&&        
-3 \big(
        17+60 \eta -135 \eta ^2+2 \eta ^3\big)
+N^2 \big(
        254+1125 \eta +180 \eta ^2+45 \eta ^3\big)
\\
P_{20}&=&50 N^2 \eta 
-660 N^3 \eta 
-440 N^4 \eta 
-3 \big(
        3+160 \eta +3 \eta ^2\big)
+9 N \big(
        7+30 \eta +7 \eta ^2\big)
\\
P_{21}&=&(2+N) \Big[
        -24 \eta 
        +48 N \eta 
        +18 N^3 \eta 
        -5 N^2 (1+9 \eta )
\Big]\\
P_{22}&=&60 N^2 (1+N)^2 \eta\\
P_{23}&=&\big(
        3-21 N+25 N^2+50 N^3+25 N^4\big) (-1+\eta ) (1+\eta )\\
P_{24}&=&-36 N^3 (-1+\eta ) \eta 
+36 N^4 \eta ^2
+N (1+15 \eta ) (-5+21 \eta )
-4 \big(
        5+27 \eta ^2\big)
\nonumber\\&&
-N^2 \big(
        5+18 \eta +189 \eta ^2\big)
\\
P_{25}&=&384 \eta 
-1664 N \eta 
+18 N^3 \big(
        4-93 \eta +4 \eta ^2\big)
+9 N^4 \big(
        4-93 \eta +4 \eta ^2\big)
\nonumber\\&&        
+N^2 \big(
        36-2501 \eta +36 \eta ^2\big)
\\
P_{26}&=&-48 \big(
        1+160 \eta +\eta ^2\big)
-16 N \big(
        4-1405 \eta +4 \eta ^2\big)
+3 N^4 \big(
        71+134 \eta +71 \eta ^2\big)
\nonumber\\&&        
+N^2 \big(
        101+15234 \eta +101 \eta ^2\big)
+2 N^3 \big(
        357+418 \eta +357 \eta ^2\big)
\\
P_{27}&=&56 N^4 (1+\eta ) \big(
        1+\eta ^2\big)
-6 (1+\eta ) \big(
        5+86 \eta +5 \eta ^2\big)
-N (1+\eta ) \big(
        41+390 \eta +41 \eta ^2\big)
\nonumber\\&&        
+4 N^3 (1+\eta ) \big(
        41+405 \eta +41 \eta ^2\big)
+2 N^2 (1+\eta ) \big(
        53+972 \eta +53 \eta ^2\big)
\\
P_{28}&=&3 \big(
        17+62 \eta -270 \eta ^2+62 \eta ^3+17 \eta ^4\big)
+20 N^4 \big(
        10+81 \eta +81 \eta ^3+10 \eta ^4\big)
\nonumber\\&&        
-N \big(
        41+768 \eta -1530 \eta ^2+768 \eta ^3+41 \eta ^4\big)
\nonumber\\&&        
+4 N^3 \big(
        41+855 \eta +270 \eta ^2+855 \eta ^3+41 \eta ^4\big)
\nonumber\\&&        
-2 N^2 \big(
        127+585 \eta +180 \eta ^2+585 \eta ^3+127 \eta ^4\big)
\\
P_{29}&=&144-51 N-585 N^2+190 N^3+36 N^4+56 N^5\\
P_{30}&=&-800 N^6
-8 N^5 (169+270 \eta )
+4 N^4 \big(
        599-645 \eta +30 \eta ^2\big)
-2 N^3 \big(
        -1565-3810 \eta 
\nonumber\\&&
        +225 \eta ^2+6 \eta ^3\big)
+3 N^2 \big(
        -355+2255 \eta +335 \eta ^2+53 \eta ^3\big)
\nonumber\\&&        
+3 \big(
        43+705 \eta +405 \eta ^2+175 \eta ^3\big)
-2 N \big(
        349+4530 \eta +855 \eta ^2+342 \eta ^3\big)
\\
P_{31}&=&-400 N^6
+24 \eta  (80+3 \eta )
-4 N^5 (119+128 \eta )
+N^3 \big(
        847+2314 \eta -513 \eta ^2\big)
\nonumber\\&&        
+N \big(
        129-4882 \eta -171 \eta ^2\big)
-4 N^4 \big(
        -359+172 \eta +3 \eta ^2\big)
\nonumber\\&&        
+4 N^2 \big(
        -239+572 \eta +189 \eta ^2\big)
\\
P_{32}&=&-280 N^6
-100 N^5 (11+81 \eta )
+N^3 \big(
        13967+41445 \eta -1755 \eta ^2-2325 \eta ^3\big)
\nonumber\\&&        
+27 \big(
        -61-160 \eta -45 \eta ^2+2 \eta ^3\big)
-90 N^4 \big(
        -65-72 \eta +27 \eta ^2+20 \eta ^3\big)
\nonumber\\&&        
+N^2 \big(
        5143+15660 \eta -405 \eta ^2+174 \eta ^3\big)
\nonumber\\&&        
+3 N \big(
        -1141-3285 \eta +1125 \eta ^2+377 \eta ^3\big)
\\
P_{33}&=&2 \big(
        -42-29 N+68 N^2+47 N^3+88 N^4+99 N^5+33 N^6\big)\\
P_{34}&=&528-224 N+2008 N^2+7149 N^3+4239 N^4+279 N^5+93 N^6\\
P_{35}&=&400 N^6
-24 \eta  (80+3 \eta )
+4 N^5 (119+128 \eta )
+4 N^4 \big(
        -359+172 \eta +3 \eta ^2\big)
\nonumber\\&&        
+N \big(
        -129+4882 \eta +171 \eta ^2\big)
-4 N^2 \big(
        -239+572 \eta +189 \eta ^2\big)
\nonumber\\&&        
+N^3 \big(
        -847-2314 \eta +513 \eta ^2\big)
\\
P_{36}&=&-1088-800 N+1664 N^2+1081 N^3+1675 N^4+1899 N^5+633 N^6\\
P_{37}&=&800 N^6
+8 N^5 (169+270 \eta )
-4 N^4 \big(
        599-645 \eta +30 \eta ^2\big)
+2 N^3 \big(
        -1565-3810 \eta
\nonumber\\&&
         +225 \eta ^2+6 \eta ^3\big)
-3 N^2 \big(
        -355+2255 \eta +335 \eta ^2+53 \eta ^3\big)
\nonumber\\&&        
-3 \big(
        43+705 \eta +405 \eta ^2+175 \eta ^3\big)
+2 N \big(
        349+4530 \eta +855 \eta ^2+342 \eta ^3\big)
\\
P_{38}&=&-400 N^6 \eta ^2
+24 (3+80 \eta )
-4 N^5 \eta  (128+119 \eta )
+N \big(
        -171-4882 \eta +129 \eta ^2\big)
\nonumber\\&&        
-4 N^2 \big(
        -189-572 \eta +239 \eta ^2\big)
+4 N^4 \big(
        -3-172 \eta +359 \eta ^2\big)
\nonumber\\&&        
+N^3 \big(
        -513+2314 \eta +847 \eta ^2\big)
\\
P_{39}&=&-512 \eta 
-256 N \eta 
+1024 N^2 \eta 
+3 N^3 \big(
        5+282 \eta +5 \eta ^2\big)
+9 N^5 \big(
        5+282 \eta +5 \eta ^2\big)
\nonumber\\&&        
+3 N^6 \big(
        5+282 \eta +5 \eta ^2\big)
+N^4 \big(
        45+2282 \eta +45 \eta ^2\big)
\\
P_{40}&=&400 N^6 \eta ^2
-24 (3+80 \eta )
+4 N^5 \eta  (128+119 \eta )
+N^3 \big(
        513-2314 \eta -847 \eta ^2\big)
\nonumber\\&&        
+N \big(
        171+4882 \eta -129 \eta ^2\big)
+4 N^2 \big(
        -189-572 \eta +239 \eta ^2\big)
\nonumber\\&&        
-4 N^4 \big(
        -3-172 \eta +359 \eta ^2\big)
\\
P_{41}&=&400 N^6 \big(
        1-\eta +\eta ^2\big)
-3 \big(
        109+446 \eta +109 \eta ^2\big)
+3 N^2 \big(
        151-1446 \eta +151 \eta ^2\big)
\nonumber\\&&        
+4 N^5 \big(
        169+101 \eta +169 \eta ^2\big)
-2 N^4 \big(
        599-1214 \eta +599 \eta ^2\big)
\nonumber\\&&        
+N \big(
        691+4694 \eta +691 \eta ^2\big)
-N^3 \big(
        1559+2026 \eta +1559 \eta ^2\big)
\\
P_{42}&=&-800 N^6 \eta ^3
-8 N^5 \eta ^2 (270+169 \eta )
+4 N^4 \eta  \big(
        30-645 \eta +599 \eta ^2\big)
\nonumber\\&&        
+3 N^2 \big(
        53+335 \eta +2255 \eta ^2-355 \eta ^3\big)
+3 \big(
        175+405 \eta +705 \eta ^2+43 \eta ^3\big)
\nonumber\\&&        
-2 N \big(
        342+855 \eta +4530 \eta ^2+349 \eta ^3\big)
+2 N^3 \big(
        -6-225 \eta +3810 \eta ^2+1565 \eta ^3\big)
\\
P_{43}&=&-280 N^6 \eta ^3
-100 N^5 \eta ^2 (81+11 \eta )
-27 \big(
        -2+45 \eta +160 \eta ^2+61 \eta ^3\big)
\nonumber\\&&        
+90 N^4 \big(
        -20-27 \eta +72 \eta ^2+65 \eta ^3\big)
-3 N \big(
        -377-1125 \eta +3285 \eta ^2+1141 \eta ^3\big)
\nonumber\\&&        
+N^2 \big(
        174-405 \eta +15660 \eta ^2+5143 \eta ^3\big)
\nonumber\\&&        
+N^3 \big(
        -2325-1755 \eta +41445 \eta ^2+13967 \eta ^3\big)
\\
P_{44}&=&800 N^6 \eta ^3
+8 N^5 \eta ^2 (270+169 \eta )
-4 N^4 \eta  \big(
        30-645 \eta +599 \eta ^2\big)
\nonumber\\&&        
-3 \big(
        175+405 \eta +705 \eta ^2+43 \eta ^3\big)
+2 N \big(
        342+855 \eta +4530 \eta ^2+349 \eta ^3\big)
\nonumber\\&&        
+3 N^2 \big(
        -53-335 \eta -2255 \eta ^2+355 \eta ^3\big)
-2 N^3 \big(
        -6-225 \eta +3810 \eta ^2+1565 \eta ^3\big)
\\
P_{45}&=&-800 N^6 \big(
        1+\eta ^4\big)
+3 \big(
        43+880 \eta +810 \eta ^2+880 \eta ^3+43 \eta ^4\big)
\nonumber\\&&        
+N^3 \big(
        3130+7608 \eta -900 \eta ^2+7608 \eta ^3+3130 \eta ^4\big)
-8 N^5 \big(
        169+270 \eta +270 \eta ^3+169 \eta ^4\big)
\nonumber\\&&        
-2 N \big(
        349+4872 \eta +1710 \eta ^2+4872 \eta ^3+349 \eta ^4\big)
\nonumber\\&&        
-3 N^2 \big(
        355-2308 \eta -670 \eta ^2-2308 \eta ^3+355 \eta ^4\big)
\nonumber\\&&        
+4 N^4 \big(
        599-645 \eta +60 \eta ^2-645 \eta ^3+599 \eta ^4\big)
\\
P_{46}&=&-800 N^6 (-1+\eta ) (1+\eta ) \big(
        1+\eta ^2\big)
-698 N (-1+\eta ) (1+\eta ) \big(
        1+12 \eta +\eta ^2\big)
\nonumber\\&&        
+3 (-1+\eta ) (1+\eta ) \big(
        43+530 \eta +43 \eta ^2\big)
-8 N^5 (-1+\eta ) (1+\eta ) \big(
        169+270 \eta +169 \eta ^2\big)
\nonumber\\&&        
-3 N^2 (-1+\eta ) (1+\eta ) \big(
        355-2202 \eta +355 \eta ^2\big)
\nonumber\\&&        
+4 N^4 (-1+\eta ) (1+\eta ) \big(
        599-645 \eta +599 \eta ^2\big)
\nonumber\\&&        
+2 N^3 (-1+\eta ) (1+\eta ) \big(
        1565+3816 \eta +1565 \eta ^2\big)
\\
P_{47}&=&800 N^6 \big(
        1+\eta ^4\big)
-3 \big(
        43+880 \eta +810 \eta ^2+880 \eta ^3+43 \eta ^4\big)
\nonumber\\&&        
+N \big(
        698+9744 \eta +3420 \eta ^2+9744 \eta ^3+698 \eta ^4\big)
+8 N^5 \big(
        169+270 \eta +270 \eta ^3+169 \eta ^4\big)
\nonumber\\&&        
+3 N^2 \big(
        355-2308 \eta -670 \eta ^2-2308 \eta ^3+355 \eta ^4\big)
\nonumber\\&&        
-4 N^4 \big(
        599-645 \eta +60 \eta ^2-645 \eta ^3+599 \eta ^4\big)
\nonumber\\&&        
-2 N^3 \big(
        1565+3804 \eta -450 \eta ^2+3804 \eta ^3+1565 \eta ^4\big)
\\
P_{48}&=&-560 N^7
-40 N^6 (62+405 \eta )
+20 N^5 \big(
        -190-2187 \eta +729 \eta ^2\big)
\nonumber\\&&        
+N^2 \big(
        1417+12690 \eta -5535 \eta ^2-240 \eta ^3\big)
-9 \big(
        17+60 \eta -135 \eta ^2+2 \eta ^3\big)
\nonumber\\&&        
+10 N^4 \big(
        -260-3807 \eta +2430 \eta ^2+15 \eta ^3\big)
-N^3 \big(
        107+9045 \eta -4455 \eta ^2+75 \eta ^3\big)
\nonumber\\&&        
-3 N \big(
        -91-1665 \eta +1575 \eta ^2+137 \eta ^3\big)
\\
P_{49}&=&-(2+N) \Big[
        -216 \eta 
        -144 N \eta 
        -696 N^2 \eta 
        +148 N^6 \eta 
        -30 N^4 \big(
                9-16 \eta +9 \eta ^2\big)
\nonumber\\&&                
        -9 N^5 \big(
                15-38 \eta +15 \eta ^2\big)
        -N^3 \big(
                135+1022 \eta +135 \eta ^2\big)
\Big]\\
P_{50}&=&225
+1630 \eta 
+3456 \eta ^2
+2466 \eta ^3
+415 \eta ^4
\nonumber\\&&
+N \big(
        930+5372 \eta +6912 \eta ^2+2820 \eta ^3+350 \eta ^4\big)
\nonumber\\&&        
-152 N^6 \eta  (1+\eta ) \big(
        5+22 \eta +5 \eta ^2\big)
-16 N^7 \eta  (1+\eta ) \big(
        5+22 \eta +5 \eta ^2\big)
\nonumber\\&&        
-192 N^3 \big(
        -5+3 \eta +135 \eta ^2+157 \eta ^3+30 \eta ^4\big)
\nonumber\\&&        
-72 N^4 \big(
        -5+53 \eta +405 \eta ^2+427 \eta ^3+80 \eta ^4\big)
\nonumber\\&&        
-12 N^5 \big(
        -5+218 \eta +1296 \eta ^2+1318 \eta ^3+245 \eta ^4\big)
-6 N^2 \big(
        -225-830 \eta 
\nonumber\\&&        
        +864 \eta ^2+1854 \eta ^3+385 \eta ^4\big)
\\
P_{51}&=&396+690 N+518 N^2+240 N^3-289 N^4-432 N^5+494 N^6+588 
N^7+147 N^8\\
P_{52}&=&560 N^8
-216 (-1+\eta ) \eta  (80+3 \eta )
+40 N^7 (34+405 \eta )
+N^4 \big(
        44539+73899 \eta
\nonumber\\&&        
         +1629 \eta ^2-7215 \eta ^3\big)
+60 N^6 \big(
        -310-561 \eta +81 \eta ^2+60 \eta ^3\big)
\nonumber\\&&        
+9 N \big(
        -420-6451 \eta +4306 \eta ^2+189 \eta ^3\big)
-2 N^5 \big(
        7334+31887 \eta -414 \eta ^2+375 \eta ^3\big)
\nonumber\\&&        
-9 N^2 \big(
        1731-919 \eta +137 \eta ^2+391 \eta ^3\big)
+3 N^3 \big(
        9966+26631 \eta -11136 \eta ^2+959 \eta ^3\big)
\\
P_{53}&=&-1080 N^2 (1+N)^2 \eta ^2 \big(
        207-456 N+89 N^2-56 N^3+92 N^4\big)\\
P_{54}&=&-540 N (1+N)^2 \eta ^2 \big(
        144-51 N-585 N^2+190 N^3+36 N^4+56 N^5\big)\\
P_{55}&=&N^2 (1+N) \Big[
        288 N^5 \eta ^2
        -36 N^4 \eta  (-8+3 \eta )
        -160 \big(
                1+3 \eta ^2\big)
        -4 N^3 \big(
                -5-9 \eta +438 \eta ^2\big)
\nonumber\\&&                
        +N \big(
                -275-270 \eta +801 \eta ^2\big)
        +N^2 \big(
                25-522 \eta +1485 \eta ^2\big)
\Big]\\
P_{56}&=&560 N^8 \eta ^3
+40 N^7 \eta ^2 (405+34 \eta )
+216 (-1+\eta ) (3+80 \eta )
-60 N^6 \big(
        -60
\nonumber\\&&        
        -81 \eta +561 \eta ^2+310 \eta ^3\big)
-9 N \big(
        -189-4306 \eta +6451 \eta ^2+420 \eta ^3\big)
\nonumber\\&&        
-9 N^2 \big(
        391+137 \eta -919 \eta ^2+1731 \eta ^3\big)
-2 N^5 \big(
        375-414 \eta +31887 \eta ^2+7334 \eta ^3\big)
\nonumber\\&&        
+3 N^3 \big(
        959-11136 \eta +26631 \eta ^2+9966 \eta ^3\big)
\nonumber\\&&        
+N^4 \big(
        -7215+1629 \eta +73899 \eta ^2+44539 \eta ^3\big)
\\
P_{57}&=&-51840 \eta ^2
+103680 N \eta ^2
+80 N^8 \eta  \big(
        405-10412 \eta +405 \eta ^2\big)
\nonumber\\&&        
-4 N^7 \big(
        -2700-20331 \eta +165688 \eta ^2+1539 \eta ^3\big)
\nonumber\\&&        
-36 N^6 \big(
        204+378 \eta -49156 \eta ^2+1863 \eta ^3\big)
\nonumber\\&&        
-9 N^2 \big(
        459+5184 \eta +20987 \eta ^2+3618 \eta ^3\big)
\nonumber\\&&        
-2 N^5 \big(
        13500+171477 \eta -942766 \eta ^2+3807 \eta ^3\big)
\nonumber\\&&        
-3 N^3 \big(
        -2025-24489 \eta +185329 \eta ^2+4077 \eta ^3\big)
\nonumber\\&&        
+N^4 \big(
        18360+94851 \eta +70490 \eta ^2+58239 \eta ^3\big)
\\
P_{58}&=&N \Big[
        2240 N^9 (-1+\eta ) (1+\eta ) \big(
                1+\eta ^2\big)
        +N^4 \big(
                6264+156159 \eta +63990 \eta ^2
\nonumber\\&&                
                -251679 \eta ^3-13850 \
\eta ^4\big)
        +N^3 \big(
                -6175-104082 \eta +24435 \eta ^2+51822 \eta ^3+3020 
\eta ^4\big)
\nonumber\\&&
        +N^2 \big(
                -525-58899 \eta -7155 \eta ^2+87081 \eta ^3+4230 \eta 
^4\big)
        +27 \big(
                -17-96 \eta +135 \eta ^2
\nonumber\\&&                
                +34 \eta ^3\big)
        +160 N^8 \big(
                -48-405 \eta +405 \eta ^3+55 \eta ^4\big)
\nonumber\\&&                
        +8 N^7 \big(
                -450-13887 \eta +17937 \eta ^3+860 \eta ^4\big)
\nonumber\\&&                
        +9 N \big(
                25-482 \eta -1845 \eta ^2+2134 \eta ^3+100 \eta ^4\big)
\nonumber\\&&                
        -10 N^5 \big(
                -2214-30615 \eta +486 \eta ^2+36153 \eta ^3+2834 \eta 
^4\big)
\nonumber\\&&
        -4 N^6 \big(
                -3660-21279 \eta +7290 \eta ^2+11559 \eta ^3+3620 
\eta ^4\big)
\Big]\\
P_{59}&=&36288 \eta 
-19872 N \eta 
-220032 N^2 \eta 
-252192 N^3 \eta 
+N^5 \big(
        -61155+394298 \eta -61155 \eta ^2\big)
\nonumber\\&&        
+N^6 \big(
        -17415+597938 \eta -17415 \eta ^2\big)
-320 N^4 \big(
        81+526 \eta +81 \eta ^2\big)
\nonumber\\&&        
+36 N^{11} \big(
        405-3766 \eta +405 \eta ^2\big)
+12 N^{12} \big(
        405-3766 \eta +405 \eta ^2\big)
\nonumber\\&&        
-5 N^9 \big(
        3483-56554 \eta +3483 \eta ^2\big)
+N^{10} \big(
        3645+44314 \eta +3645 \eta ^2\big)
\nonumber\\&&        
+10 N^7 \big(
        4455-2386 \eta +4455 \eta ^2\big)
+2 N^8 \big(
        7695-64514 \eta +7695 \eta ^2\big)
\\
P_{60}&=&1080 N^3 (1+N)^2 (2+N)^4 \eta ^3 \big(
        576+1173 N-2988 N^2-4835 N^3+2674 N^4+572 N^5
\nonumber\\&&
        +248 N^6\big)\\
P_{61}&=&-5806080 \eta ^3
-7464960 N \eta ^3
+39628800 N^2 \eta ^3
\nonumber\\&&
+N^4 \big(
        273375-2809566 \eta +5935680 \eta ^2+522717086 \eta 
^3+2240865 \eta ^4-4790016 \eta ^5\big)
\nonumber\\&&
+864 N^{15} \eta  \big(
        400+3639 \eta +589 \eta ^2+3639 \eta ^3+400 \eta ^4\big)
\nonumber\\&&        
-768 N^3 \eta  \big(
        1377+4293 \eta -244034 \eta ^2+4293 \eta ^3+1377 \eta ^4\big)
\nonumber\\&&        
+432 N^{14} \eta  \big(
        7856+80679 \eta +16689 \eta ^2+80679 \eta ^3+7856 \eta 
^4\big)
\nonumber\\&&
-16 N^{11} \eta  \big(
        450738-6962355 \eta -48003782 \eta ^2-6819795 \eta ^3+483138 
\eta ^4\big)
\nonumber\\&&
+8 N^{13} \eta  \big(
        1570752+18898191 \eta +7445585 \eta ^2+18898191 \eta 
^3+1570752 \eta ^4\big)
\nonumber\\&&
+4 N^{12} \eta  \big(
        4566888+73693719 \eta +72807653 \eta ^2+73622439 \eta 
^3+4550688 \eta ^4\big)
\nonumber\\&&
+4 N^5 \big(
        236925+975942 \eta +27088992 \eta ^2+117199930 \eta 
^3+25711587 \eta ^4-325728 \eta ^5\big)
\nonumber\\&&
-12 N^8 \big(
        28350+912078 \eta +59340789 \eta ^2+250807007 \eta 
^3+58332339 \eta ^4+564588 \eta ^5\big)
\nonumber\\&&
-36 N^9 \big(
        6075+1870314 \eta +30831858 \eta ^2+19135996 \eta ^3+30932703 
\eta ^4+1867884 \eta ^5\big)
\nonumber\\&&
-8 N^{10} \big(
        6075+7288137 \eta +73658484 \eta ^2-103483664 \eta 
^3+74427579 \eta ^4+7437582 \eta ^5\big)
\nonumber\\&&
+4 N^7 \big(
        18225+8623746 \eta +9768924 \eta ^2-816997534 \eta 
^3+15310539 \eta ^4+9807156 \eta ^5\big)
\nonumber\\&&
+2 N^6 \big(
        443475+12987162 \eta +132854256 \eta ^2-575608138 \eta 
^3+137257821 \eta ^4
\nonumber\\&&
+12137472 \eta ^5\big).
\end{eqnarray}


In the expression of $\tilde{a}_{gg,Q}^{(3)}$, denominators with poles at $N = 1/2$ and $N = 3/2$ contribute.
We have checked using the algorithms of {\tt HarmonicSums} that there are no singularities at these points.
The proof of this can either be performed in Mellin $N$ space or in $z$ space. Even using the technologies 
available in {\tt HarmonicSums} it is far from being trivial, since various new iterative integrals depending on $\eta$
emerge, the cancellation of which have to be shown analytically. In general, it is necessary to reduce these integrals 
to higher functions known, cf.~\cite{Blumlein:2018cms}. This is not always simple because of a proliferation of letters
for the corresponding integrals, see (\ref{eq:GG}). In the expansion around $N = 1/2,~3/2$ letters occur, which do not belong
to the class of those emerging in the $N$- or $z$ space results. Examples for the associated iterative integrals are 
\begin{eqnarray}
&& G\left(\left\{-\frac{\sqrt{2} - \sqrt{z(1+z)}}{1-z}, \frac{\sqrt{1-z^2}}{z},\frac{1}{z}\right\};1\right)
\\
&& G\left(\left\{-\frac{(1-\sqrt{(2-z)z}}{1-z}, \sqrt{(1-z)(2-z)}, \frac{1}{1-z}\right\};1\right)
\\
&& G\left(\left\{\frac{z}{2-z^2}, \frac{1}{1+z}, \frac{1}{1-z}\right\};1\right)
\\
&& G\left(\left\{\frac{\sqrt{1-z^2}}{z},\frac{1}{1-z},-\frac{1}{\sqrt{2} + \sqrt{1+z}}\right\};1\right).
\end{eqnarray}
Fortunately, these functions cancel in the physical expressions and have not to be simplified.

For $N = 3,5$ we can compare to the fixed moments calculated in Section~\ref{sec:3} and find agreement for the terms 
within the accuracy the $\eta$ expansion has been performed to. 
\section{The Transformation to \boldmath $z$ Space} 
\label{sec:5} 

\vspace*{1mm}
\noindent
From the Mellin space result we compute the inverse Mellin transform to $z$ space using the methods of 
\cite{Ablinger:2016lzr,Ablinger:2018cja,Ablinger:2018pwq}, which are implemented in the package \texttt{HarmonicSums}. 
The idea is to find the difference equation satisfied by the sums appearing in $N$ space, and convert 
them into 
differential equations to be satisfied by the inverse Mellin transform, and to solve them. The result is then given 
in the form of harmonic polylogarithms and more general iterated integrals of the type
\begin{eqnarray}
\label{eq:GG}
G\left[\left\{g(x),\vec{h}(x)\right\},z\right] &=& \int_0^z \dd y\, g(y) 
G\left[\left\{\vec{h}(x)\right\},y\right].
\end{eqnarray}
In what follows we define 
\begin{eqnarray}
n &=& N-1 \\
\Mvec\left[ g(z) \right] &=& \int_{0}^1 \dd z\, z^n g(z) = g(n) \\
\Mvec^{-1} \left[ g(n) \right] &=& g(z).
\end{eqnarray}
In general, the result of the inverse Mellin transform of a function $D$ will contain different pieces
\begin{eqnarray}
        D (z) &\propto & \frac{1-(-1)^N}{2} \bigg\{D^{\delta} 
\delta (1-z) + D^{+} (z) + D^{\text{reg}} (z) + \sum_a\Mvec^{-1} \left[ n^a g_a (n) \right] (z) \bigg\} .
\end{eqnarray}
One distinguishes between the distribution $D^\delta$, $D^+$, and 
$D^{\rm reg}$, the regular part, where $D^\delta$ 
is a function of $\eta$ and $D^+$ is a $+$-distribution,
\begin{equation}
\int_0^1 \dd z\,g(z) \big(f(z)\big)_+ = \int_0^1\dd z\,\Big(g(z)-g(1)\Big)f(z),
\end{equation}
the regular part in $z$,  $D^{\text{reg}}$. At times, it is  necessary to absorb from the output of \texttt{HarmonicSums} 
a rational pre--factor in $n$ after applying a partial fractioning of the result, using the relations
\begin{eqnarray}
\label{eq:abs1}
        \frac{1}{(n+a)^i} \int_{0}^{1} \text{d}z \ z^n \ f(z) &=& \int_{0}^{1} \text{d}z \ z^n \biggl\{ \int_z^1 
\text{d} y \ (-1)^{i-1} \left( \frac{y}{z} \right)^{a} \left[ \HA_0 \left( \frac{y}{z} \right) \right]^{i-1}  f(y)    
\biggr\} \\ \label{eq:abs2}
        n \int_{0}^{1} \text{d}z \ z^n \ f(z) &=& \left. ( z^n - 1 ) z f(z) \right|_{0}^{1} - \int_{0}^{1} (z^n - 
1) \frac{d}{dz} \left( z f(z) \right). 
\end{eqnarray}
We denote the Mellin convolution by $\otimes$ 
\begin{equation}
f(z) \otimes g(z) = \int_{0}^{1}\dd z_{1}\int_{0}^{1}\dd z_{2}\,\delta(z-z_{1}z_{2})f(z_{1})g(z_{2})
\end{equation}
for regular functions and, cf.~\cite{Ablinger:2014bra},
\begin{equation}
[f(z)]_+ \otimes g(z) = \int_{z}^{1}dy\,f(y)\left[\frac{1}{y}g\left(\frac{z}{y}\right)-g(z)\right]-g(z)\int_{0}^{z}dy\,f(y)
\end{equation}
for the Mellin convolution of a $+$-distribution and a regular function.

Now we turn to $a_{gg,Q}^{(3)}(z)$, which we represent in terms of the three contributing parts
combining to
\begin{eqnarray}
\tilde{a}_{gg,Q}^{(3)} (N) &=& \int_{0}^{1} \text{d} z \ z^{N-1} \ \delta (1-z) \ \tilde{a}_{gg,Q}^{(3),\delta} (z) + \int_{0}^{1} \text{d} z \ \left( z^{N-1} - 1 \right) \ \tilde{a}_{gg,Q}^{(3),+} (z)
\NN \\ 
 &+& \int_{0}^{1} \text{d} z \ z^{N-1} \ \tilde{a}_{gg,Q}^{(3),\text{reg}} (z) ~.
\end{eqnarray}
The result is expressed in terms of the iterated integrals $G_i$ and the constants $K_i$ defined in 
Appendix~D of \cite{Ablinger:2018brx}.
In the following the argument of the functions $G_i$ is implied to be $z$ in the formulas for 
$\tilde{a}_{gg,Q}^{(3),\delta} (z)$, $\tilde{a}_{gg,Q}^{(3),+} (z)$ and 
$\tilde{a}_{gg,Q}^{(3),\text{reg}}(z)$, and it is implied to be $y$ in the functions $\Phi_i$ which 
follow. Such arguments are omitted in the interest of brevity. One finds


with the polynomials
\begin{eqnarray}
Q_1&=&-405
-405 \eta 
+10412 \eta ^2
-405 \eta ^3
-405 \eta ^4
+405 z (-1+\eta )^2 \big(
        1+\eta +\eta ^2\big)
\\
Q_2&=&-z (-1+\eta )^2 \big(
        1+\eta ^4\big)
-2 \eta  \big(
        1+\eta ^4\big)
+z^2 (-1+\eta )^2 (1+\eta )^2 \big(
        1-\eta +\eta ^2\big)
\\
Q_3&=&2 \eta  \big(
        1-\eta +\eta ^2\big)
+z^3 (-1+\eta )^2 \big(
        1+\eta +\eta ^2\big)
+z \big(
        1-6 \eta +6 \eta ^2-6 \eta ^3+\eta ^4\big)
\nonumber\\&&        
-z^2 (-1+\eta )^2 \big(
        2-\eta +2 \eta ^2\big)
\\
Q_4&=&30 \eta 
+88 \eta ^3
+30 \eta ^5
+z \big(
        15-60 \eta +103 \eta ^2-176 \eta ^3+103 \eta ^4-60 \eta ^5+15 
\eta ^6\big)
\nonumber\\&&
+15 z^3 (-1+\eta )^2 (1+\eta )^2 \big(
        1-\eta +\eta ^2\big)
\nonumber\\&&        
-z^2 (-1+\eta )^2 \big(
        30+15 \eta +88 \eta ^2+15 \eta ^3+30 \eta ^4\big)
\\
Q_5&=&5
+22 \eta 
+5 \eta ^2
+2 z \big(
        1-10 \eta +\eta ^2\big)
\\
Q_6&=&45
+302 \eta 
+45 \eta ^2
+z \big(
        27-10 \eta +27 \eta ^2\big)
\\
Q_7&=&135
-3436 \eta 
+135 \eta ^2
+z \big(
        81+596 \eta +81 \eta ^2\big)
\\
Q_8&=&z^2 \big(
        -287+62 \eta -287 \eta ^2\big)
+120 z^4 \big(
        1-10 \eta +\eta ^2\big)
-7 \big(
        1-\eta +\eta ^2\big)
\nonumber\\&&        
+240 z^3 \big(
        1+8 \eta +\eta ^2\big)
-6 z \big(
        11+64 \eta +11 \eta ^2\big)
\\
Q_9&=&16 \big(
        73+90 \eta +163 \eta ^2+90 \eta ^3+73 \eta ^4\big)\\
Q_{10}&=&20 (1+\eta )^2 \big(
        73+17 \eta +73 \eta ^2\big)\\
Q_{11}&=&1
-70 \eta 
+\eta ^2
+8 z \big(
        1+40 \eta +\eta ^2\big)
\\
Q_{12}&=&29
+2540 \eta 
+29 \eta ^2
+2 z \big(
        16-1523 \eta +16 \eta ^2\big)
\\
Q_{13}&=&9
+400 \eta 
+9 \eta ^2
+4 z \big(
        9+100 \eta +9 \eta ^2\big)
\\
Q_{14}&=&109
+446 \eta 
+109 \eta ^2
+64 z \big(
        1+14 \eta +\eta ^2\big)
\\
Q_{15}&=&9
+160 \eta 
+9 \eta ^2
+8 z \big(
        9+20 \eta +9 \eta ^2\big)
\\
Q_{16}&=&-81
-410 \eta 
-81 \eta ^2
+z \big(
        81+550 \eta +81 \eta ^2\big)
\\
Q_{17}&=&-81
-820 \eta 
-81 \eta ^2
+3 z \big(
        27+260 \eta +27 \eta ^2\big)
\\
Q_{18}&=&59
+226 \eta 
+59 \eta ^2
+4 z \big(
        59+346 \eta +59 \eta ^2\big)
\\
Q_{19}&=&9
-11246 \eta 
+9 \eta ^2
+8 z \big(
        261+1568 \eta +261 \eta ^2\big)
\\
Q_{20}&=&-5 \big(
        17739+24192 \eta +397052 \eta ^2+24192 \eta ^3+17739 \eta 
^4\big)
\nonumber\\&&
+z \big(
        88695+567 \eta +2287160 \eta ^2+567 \eta ^3+88695 \eta 
^4\big)
\\
Q_{21}&=&-49 \eta  \Big[
        -z (-1+\eta )^2
        +z^2 (-1+\eta )^2
        -\eta 
\Big]\\
Q_{22}&=&147 \eta  \big(
        1+\eta ^2
\big)
\Big[-z (-1+\eta )^2
        +z^2 (-1+\eta )^2
        -\eta 
\Big]\\
Q_{23}&=&5 \eta  \big(
        807+574 \eta +807 \eta ^2\big)
+z \big(
        3285-6985 \eta -7249 \eta ^2-6985 \eta ^3+3285 \eta ^4\big)
\nonumber\\&&        
+3 z^3 (-1+\eta )^2 \big(
        1095+2693 \eta +1095 \eta ^2\big)
\nonumber\\&&        
-z^2 (-1+\eta )^2 \big(
        6570+11699 \eta +6570 \eta ^2\big)
\\
Q_{24}&=&\eta  \big(
        2421+792 \eta -33824 \eta ^2+792 \eta ^3+2421 \eta ^4\big)
\nonumber\\&&        
+z \big(
        1971-5121 \eta -40574 \eta ^2+67548 \eta ^3-40574 \eta 
^4-5121 \eta ^5+1971 \eta ^6\big)
\nonumber\\&&
+z^3 (-1+\eta )^2 \big(
        1971+7137 \eta +1684 \eta ^2+7137 \eta ^3+1971 \eta ^4\big)
\nonumber\\&&        
-z^2 (-1+\eta )^2 \big(
        3942+8379 \eta -32140 \eta ^2+8379 \eta ^3+3942 \eta ^4\big)
\\
Q_{25}&=&z \big(
        9855-4695 \eta -27488 \eta ^2-123060 \eta ^3-27488 \eta 
^4-4695 \eta ^5+9855 \eta ^6\big)
\nonumber\\&&
+5 \eta  \big(
        2421+4974 \eta +8324 \eta ^2+4974 \eta ^3+2421 \eta ^4\big)
\nonumber\\&&        
+z^3 (-1+\eta )^2 \big(
        9855+29223 \eta +89560 \eta ^2+29223 \eta ^3+9855 \eta 
^4\big)
\nonumber\\&&
-z^2 (-1+\eta )^2 \big(
        19710+56343 \eta +131180 \eta ^2+56343 \eta ^3+19710 \eta 
^4\big)
\\
Q_{26}&=&49 \big(
        1-\eta +\eta ^2\big)
+3840 z^5 \big(
        1+14 \eta +\eta ^2\big)
+60 z^4 \big(
        13-898 \eta +13 \eta ^2\big)
\nonumber\\&&        
-58 z^3 \big(
        173+1102 \eta +173 \eta ^2\big)
+z \big(
        1463+412 \eta +1463 \eta ^2\big)
\nonumber\\&&        
+z^2 \big(
        7187+64438 \eta +7187 \eta ^2\big).
\end{eqnarray}

The functions $\Phi_1,\ldots,\Phi_8$, which appear as arguments of a further integral, are


where the polynomials $R_i$ are:
\begin{eqnarray}
R_1&=&-20 (1+\eta )^2 \big(
        1-10 \eta +\eta ^2\big)\\
R_2&=&-16 \big(
        1-9 \eta -8 \eta ^2-9 \eta ^3+\eta ^4\big)\\
R_3&=&\eta  \big(
        729-862 \eta +729 \eta ^2\big)
+27 y^2 (-1+\eta )^2 \big(
        1+46 \eta +\eta ^2\big)
\nonumber\\&&        
-27 y \big(
        1+70 \eta -126 \eta ^2+70 \eta ^3+\eta ^4\big)
\\
R_4&=&18 \eta ^2
+2 y \eta  \big(
        13-50 \eta +13 \eta ^2\big)
+y^3 (-1+\eta )^2 \big(
        1+46 \eta +\eta ^2\big)
\nonumber\\&&        
-y^2 (-1+\eta )^2 \big(
        1+74 \eta +\eta ^2\big)
\\
R_5&=&2 \eta ^2 \big(
        81-34 \eta +81 \eta ^2\big)
\nonumber\\&&        
+y^2 \big(
        9-576 \eta +1391 \eta ^2-1360 \eta ^3+1391 \eta ^4-576 \eta 
^5+9 \eta ^6\big)
\nonumber\\&&
+9 y^4 (-1+\eta )^2 (1+\eta )^2 \big(
        1-10 \eta +\eta ^2\big)
-18 y^3 (-1+\eta )^2 \big(
        1-21 \eta -21 \eta ^3+\eta ^4\big)
\nonumber\\&&        
+2 y \eta  \big(
        126-385 \eta +302 \eta ^2-385 \eta ^3+126 \eta ^4\big)
\\
R_6&=&2 \eta ^2 \big(
        81-62 \eta +81 \eta ^2\big)
\nonumber\\&&        
+y^2 \big(
        9-576 \eta +1447 \eta ^2-1472 \eta ^3+1447 \eta ^4-576 \eta 
^5+9 \eta ^6\big)
\nonumber\\&&
+9 y^4 (-1+\eta )^2 (1+\eta )^2 \big(
        1-10 \eta +\eta ^2\big)
-18 y^3 (-1+\eta )^2 \big(
        1-21 \eta -21 \eta ^3+\eta ^4\big)
\nonumber\\&&        
+2 y \eta  \big(
        126-413 \eta +358 \eta ^2-413 \eta ^3+126 \eta ^4\big)
\\
R_7&=&-20 (1+\eta )^2 \big(
        5+22 \eta +5 \eta ^2\big)\\
R_8&=&-16 \big(
        5+27 \eta +32 \eta ^2+27 \eta ^3+5 \eta ^4\big)\\
R_9&=&2 \eta  \big(
        11-70 \eta +11 \eta ^2\big)
+y^2 (-1+\eta )^2 \big(
        5+86 \eta +5 \eta ^2\big)
\nonumber\\&&        
-y (-1+\eta )^2 \big(
        5+118 \eta +5 \eta ^2\big)
\\
R_{10}&=&\eta  \big(
        243-790 \eta +243 \eta ^2\big)
+9 y^2 (-1+\eta )^2 \big(
        5+86 \eta +5 \eta ^2\big)
\nonumber\\&&        
-9 y \big(
        5+98 \eta -270 \eta ^2+98 \eta ^3+5 \eta ^4\big)
\\
R_{11}&=&y \Big[
        -280 \eta ^3
        +y^2 \big(
                15-96 \eta +757 \eta ^2-1736 \eta ^3+757 \eta ^4-96 
\eta ^5+15 \eta ^6\big)
\nonumber\\&&
        +3 y^4 (-1+\eta )^2 (1+\eta )^2 \big(
                5+22 \eta +5 \eta ^2\big)
\nonumber\\&&                
        -6 y^3 (-1+\eta )^2 \big(
                5+21 \eta +108 \eta ^2+21 \eta ^3+5 \eta ^4\big)
\nonumber\\&&                
        +4 y \eta  \big(
                24-79 \eta +254 \eta ^2-79 \eta ^3+24 \eta ^4\big)
\Big]\\
R_{12}&=&y \Big[
        -368 \eta ^3
        +y^2 \big(
                15-96 \eta +845 \eta ^2-1912 \eta ^3+845 \eta ^4-96 
\eta ^5+15 \eta ^6\big)
\nonumber\\&&
        +3 y^4 (-1+\eta )^2 (1+\eta )^2 \big(
                5+22 \eta +5 \eta ^2\big)
\nonumber\\&&                
        -6 y^3 (-1+\eta )^2 \big(
                5+21 \eta +108 \eta ^2+21 \eta ^3+5 \eta ^4\big)
\nonumber\\&&                
        +4 y \eta  \big(
                24-101 \eta +298 \eta ^2-101 \eta ^3+24 \eta ^4\big)
\Big]\\
R_{13}&=&-20 (1+\eta )^2 \big(
        29-74 \eta +29 \eta ^2\big)\\
R_{14}&=&16 \big(
        -29+45 \eta +16 \eta ^2+45 \eta ^3-29 \eta ^4\big)\\
R_{15}&=&-55 \eta  \big(
        243-382 \eta +243 \eta ^2\big)
+y \big(
        783+23949 \eta -54320 \eta ^2+23949 \eta ^3+783 \eta ^4\big)
\\
R_{16}&=&(1+\eta ) \Big[
        270 \eta ^2
        +2 y \eta  \big(
                233-982 \eta +233 \eta ^2\big)
        +y^3 (-1+\eta )^2 \big(
                29+974 \eta +29 \eta ^2\big)
\nonumber\\&&                
        -y^2 (-1+\eta )^2 \big(
                29+1498 \eta +29 \eta ^2\big)
\Big]\\
R_{17}&=&-2 \eta ^2 \big(
        1215-2686 \eta +1215 \eta ^2\big)
\nonumber\\&&        
-9 y^2 (-1+\eta )^2 \big(
        29-482 \eta +810 \eta ^2-482 \eta ^3+29 \eta ^4\big)
\nonumber\\&&        
+y^3 (-1+\eta )^2 \big(
        261-144 \eta -1610 \eta ^2-144 \eta ^3+261 \eta ^4\big)
\nonumber\\&&        
-4 y \eta  \big(
        1179-3476 \eta +4250 \eta ^2-3476 \eta ^3+1179 \eta ^4\big)
\\
R_{18}&=&-2 \eta ^2 \big(
        1215-2486 \eta +1215 \eta ^2\big)
\nonumber\\&&        
-9 y^2 (-1+\eta )^2 \big(
        29-482 \eta +810 \eta ^2-482 \eta ^3+29 \eta ^4\big)
\nonumber\\&&        
+y^3 (-1+\eta )^2 \big(
        261-144 \eta -1210 \eta ^2-144 \eta ^3+261 \eta ^4\big)
\nonumber\\&&        
-4 y \eta  \big(
        1179-3376 \eta +4150 \eta ^2-3376 \eta ^3+1179 \eta ^4\big)
\\
R_{19}&=&-20 (1+\eta )^2 \big(
        55+26 \eta +55 \eta ^2\big)\\
R_{20}&=&-16 \big(
        55+81 \eta +136 \eta ^2+81 \eta ^3+55 \eta ^4\big)\\
R_{21}&=&\eta  \big(
        -2187+4202 \eta -2187 \eta ^2\big)
+3 y (-1+\eta )^2 \big(
        55+1594 \eta +55 \eta ^2\big)
\\
R_{22}&=&(1+\eta ) \Big[
        324 \eta ^2
        +2 y \eta  \big(
                337-1526 \eta +337 \eta ^2\big)
        +y^3 (-1+\eta )^2 \big(
                55+1594 \eta +55 \eta ^2\big)
\nonumber\\&&                
        -y^2 (-1+\eta )^2 \big(
                55+2378 \eta +55 \eta ^2\big)
\Big]\\
R_{23}&=&-108 (-3+\eta ) \eta ^2 (-1+3 \eta )
+y^3 (-1+\eta )^2 (1+\eta )^2 \big(
        55+26 \eta +55 \eta ^2\big)
\nonumber\\&&        
-y^2 (-1+\eta )^2 \big(
        55-538 \eta +1942 \eta ^2-538 \eta ^3+55 \eta ^4\big)
\nonumber\\&&        
-4 y \eta  \big(
        196-573 \eta +890 \eta ^2-573 \eta ^3+196 \eta ^4\big)
\\
R_{24}&=&-4 \eta ^2 \big(
        81-250 \eta +81 \eta ^2\big)
+y^3 (-1+\eta )^2 (1+\eta )^2 \big(
        55+26 \eta +55 \eta ^2\big)
\nonumber\\&&        
-y^2 (-1+\eta )^2 \big(
        55-538 \eta +1862 \eta ^2-538 \eta ^3+55 \eta ^4\big)
\nonumber\\&&        
-4 y \eta  \big(
        196-553 \eta +850 \eta ^2-553 \eta ^3+196 \eta ^4\big)
\\
R_{25}&=&\eta 
+\eta ^3
+y \eta  \big(
        1+\eta ^2\big)
+y^2 (-1+\eta )^2 \big(
        1+\eta +\eta ^2\big)
\\
R_{26}&=&y (1+\eta ) \Big[
        -y (-1+\eta )^2 (1+\eta )^2
        +y^2 (-1+\eta )^2 \big(
                1+\eta +\eta ^2\big)
\nonumber\\&&                
        -\eta  \big(
                1+\eta +\eta ^2\big)
\Big]\\
R_{27}&=&2 \eta ^3
-y \eta  (1+\eta )^2 \big(
        1-\eta +\eta ^2\big)
+y^3 (-1+\eta )^2 (1+\eta )^2 \big(
        1-\eta +\eta ^2\big)
\nonumber\\&&
-y^2 (-1+\eta )^2 \big(
        1+2 \eta +2 \eta ^3+\eta ^4\big)
\\
R_{28}&=&-128 \big(
        1+15 \eta +16 \eta ^2+15 \eta ^3+\eta ^4\big)\\
R_{29}&=&-32 (1+\eta )^2 \big(
        1+14 \eta +\eta ^2\big)\\
R_{30}&=&189
+810 y^2 (-1+\eta )^4
-4518 \eta 
-2458 \eta ^2
-4518 \eta ^3
+189 \eta ^4
\nonumber\\&&
-27 y \big(
        37-308 \eta +30 \eta ^2-308 \eta ^3+37 \eta ^4\big)
\\
R_{31}&=&90 y^4 (-1+\eta )^4
-2 \eta  \big(
        7-158 \eta +7 \eta ^2\big)
\nonumber\\&&        
+y \big(
        -21+634 \eta -1418 \eta ^2+634 \eta ^3-21 \eta ^4\big)
\nonumber\\&&        
+4 y^2 (-1+\eta )^2 \big(
        33-296 \eta +33 \eta ^2\big)
-3 y^3 (-1+\eta )^2 \big(
        67-262 \eta +67 \eta ^2\big)
\\
R_{32}&=&y \big(
        63-1488 \eta +5429 \eta ^2+5816 \eta ^3+5429 \eta ^4-1488 
\eta ^5+63 \eta ^6\big)
\nonumber\\&&
+144 y^4 (-1+\eta )^2 (1+\eta )^2 \big(
        1+14 \eta +\eta ^2\big)
\nonumber\\&&        
-45 y^3 (-1+\eta )^2 \big(
        5+140 \eta +222 \eta ^2+140 \eta ^3+5 \eta ^4\big)
\nonumber\\&&        
+2 \eta  \big(
        21-474 \eta -470 \eta ^2-474 \eta ^3+21 \eta ^4\big)
\nonumber\\&&        
+2 y^2 \big(
        9+2640 \eta -3217 \eta ^2-3472 \eta ^3-3217 \eta ^4+2640 \eta 
^5+9 \eta ^6\big)
\\
R_{33}&=&y \big(
        63-1488 \eta +4789 \eta ^2+7096 \eta ^3+4789 \eta ^4-1488 
\eta ^5+63 \eta ^6\big)
\nonumber\\&&
+144 y^4 (-1+\eta )^2 (1+\eta )^2 \big(
        1+14 \eta +\eta ^2\big)
\nonumber\\&&        
-45 y^3 (-1+\eta )^2 \big(
        5+140 \eta +222 \eta ^2+140 \eta ^3+5 \eta ^4\big)
\nonumber\\&&        
+2 \eta  \big(
        21-474 \eta -790 \eta ^2-474 \eta ^3+21 \eta ^4\big)
\nonumber\\&&        
+2 y^2 \big(
        9+2640 \eta -2897 \eta ^2-4112 \eta ^3-2897 \eta ^4+2640 \eta 
^5+9 \eta ^6\big)
\\
R_{34}&=&-10 (1+\eta )^2 \big(
        109+446 \eta +109 \eta ^2\big)\\
R_{35}&=&-8 \big(
        109+555 \eta +664 \eta ^2+555 \eta ^3+109 \eta ^4\big)\\
R_{36}&=&-387
-5958 \eta 
-10102 \eta ^2
-5958 \eta ^3
-387 \eta ^4
\nonumber\\&&
+108 y^2 (-1+\eta )^2 \big(
        11-14 \eta +11 \eta ^2\big)
\nonumber\\&&        
-9 y \big(
        89-1312 \eta -210 \eta ^2-1312 \eta ^3+89 \eta ^4\big)
\\
R_{37}&=&43
+794 \eta 
-1530 \eta ^2
+794 \eta ^3
+43 \eta ^4
+12 y^3 (-1+\eta )^2 \big(
        11-14 \eta +11 \eta ^2\big)
\nonumber\\&&        
-y^2 (-1+\eta )^2 \big(
        221-866 \eta +221 \eta ^2\big)
\nonumber\\&&        
+2 y \big(
        23-835 \eta +1576 \eta ^2-835 \eta ^3+23 \eta ^4\big)
\\
R_{38}&=&y \Big[
        5120 \eta ^3
        +y \big(
                129+1332 \eta -4129 \eta ^2-18568 \eta ^3-4129 \eta 
^4+1332 \eta ^5+129 \eta ^6\big)
\nonumber\\&&
        -6 y^4 (-1+\eta )^2 (1+\eta )^2 \big(
                109+446 \eta +109 \eta ^2\big)
\nonumber\\&&                
        +3 y^3 (-1+\eta )^2 \big(
                479+3536 \eta +5250 \eta ^2+3536 \eta ^3+479 \eta 
^4\big)
\nonumber\\&&
        -2 y^2 \big(
                456+3195 \eta -3680 \eta ^2-7910 \eta ^3-3680 \eta 
^4+3195 \eta ^5+456 \eta ^6\big)
\Big]\\
R_{39}&=&y \Big[
        2560 \eta ^3
        +y \big(
                129+1332 \eta -6689 \eta ^2-13448 \eta ^3-6689 \eta 
^4+1332 \eta ^5+129 \eta ^6\big)
\nonumber\\&&
        -6 y^4 (-1+\eta )^2 (1+\eta )^2 \big(
                109+446 \eta +109 \eta ^2\big)
\nonumber\\&&                
        +3 y^3 (-1+\eta )^2 \big(
                479+3536 \eta +5250 \eta ^2+3536 \eta ^3+479 \eta 
^4\big)
\nonumber\\&&
        -2 y^2 \big(
                456+3195 \eta -4960 \eta ^2-5350 \eta ^3-4960 \eta 
^4+3195 \eta ^5+456 \eta ^6\big)
\Big]\\
R_{40}&=&-160 (1+\eta )^2 \big(
        17+883 \eta +17 \eta ^2\big)\\
R_{41}&=&-128 \big(
        17+900 \eta +917 \eta ^2+900 \eta ^3+17 \eta ^4\big)\\
R_{42}&=&-5 \big(
        1557-21798 \eta +3602 \eta ^2-21798 \eta ^3+1557 \eta ^4\big)
\nonumber\\&&        
+2 y \big(
        9009-53793 \eta +49720 \eta ^2-53793 \eta ^3+9009 \eta ^4\big)
\\
R_{43}&=&(1+\eta ) \Big[
        10 \eta  \big(
                53-690 \eta +53 \eta ^2\big)
\nonumber\\&&                
        +y \big(
                865-15172 \eta +30678 \eta ^2-15172 \eta ^3+865 \eta 
^4\big)
\nonumber\\&&
        +2 y^3 (-1+\eta )^2 \big(
                1001-3034 \eta +1001 \eta ^2\big)
\nonumber\\&&                
        -y^2 (-1+\eta )^2 \big(
                2867-17786 \eta +2867 \eta ^2\big)
\Big]\\
R_{44}&=&2 \eta  \big(
        -265+3450 \eta +2022 \eta ^2+3450 \eta ^3-265 \eta ^4\big)
\nonumber\\&&        
+8 y^3 (-1+\eta )^2 \big(
        34+1969 \eta +4900 \eta ^2+1969 \eta ^3+34 \eta ^4\big)
\nonumber\\&&        
+y^2 (-1+\eta )^2 \big(
        593-27584 \eta -34050 \eta ^2-27584 \eta ^3+593 \eta ^4\big)
\nonumber\\&&        
-y \big(
        865-12898 \eta +35567 \eta ^2+24180 \eta ^3+35567 \eta 
^4-12898 \eta ^5+865 \eta ^6\big)
\\
R_{45}&=&2 \eta  \big(
        -795+10350 \eta +10666 \eta ^2+10350 \eta ^3-795 \eta ^4\big)
\nonumber\\&&        
+8 y^3 (-1+\eta )^2 \big(
        102+5907 \eta +13550 \eta ^2+5907 \eta ^3+102 \eta ^4\big)
\nonumber\\&&        
+3 y^2 (-1+\eta )^2 \big(
        593-27584 \eta -34050 \eta ^2-27584 \eta ^3+593 \eta ^4\big)
\nonumber\\&&        
-y \big(
        2595-38694 \eta +97501 \eta ^2+81740 \eta ^3+97501 \eta 
^4-38694 \eta ^5+2595 \eta ^6\big)
\\
R_{46}&=&-20 (1+\eta )^2 \big(
        1127+6478 \eta +1127 \eta ^2\big)\\
R_{47}&=&-16 \big(
        1127+7605 \eta +8732 \eta ^2+7605 \eta ^3+1127 \eta ^4\big)\\
R_{48}&=&1305
+16902 \eta 
+1666 \eta ^2
+16902 \eta ^3
+1305 \eta ^4
\nonumber\\&&
+3 y (-1+\eta )^2 \big(
        257-4018 \eta +257 \eta ^2\big)
\\
R_{49}&=&(1+\eta ) \Big[
        -3 \eta  \big(
                43+1046 \eta +43 \eta ^2\big)
        +2 y^2 (-1+\eta )^2 \big(
                89+4957 \eta +89 \eta ^2\big)
\nonumber\\&&                
        +y^3 (-1+\eta )^2 \big(
                257-4018 \eta +257 \eta ^2\big)
\nonumber\\&&                
        -y \big(
                435+5633 \eta -15640 \eta ^2+5633 \eta ^3+435 \eta ^4
\big)
\Big]\\
R_{50}&=&y \big(
        -435-4249 \eta +17803 \eta ^2+8690 \eta ^3+17803 \eta ^4-4249 
\eta ^5-435 \eta ^6\big)
\nonumber\\&&
-y^3 (-1+\eta )^2 (1+\eta )^2 \big(
        1127+6478 \eta +1127 \eta ^2\big)
\nonumber\\&&        
-\eta \big(
        129+3138 \eta +3074 \eta ^2+3138 \eta ^3+129 \eta ^4\big)
\nonumber\\&&        
+2 y^2 (-1+\eta )^2 \big(
        781+7358 \eta +5990 \eta ^2+7358 \eta ^3+781 \eta ^4\big)
\\
R_{51}&=&y \big(
        -435-4249 \eta +14523 \eta ^2+15250 \eta ^3+14523 \eta 
^4-4249 \eta ^5-435 \eta ^6\big)
\nonumber\\&&
-y^3 (-1+\eta )^2 (1+\eta )^2 \big(
        1127+6478 \eta +1127 \eta ^2\big)
\nonumber\\&&        
-3 \eta \big(
        43+1046 \eta +2118 \eta ^2+1046 \eta ^3+43 \eta ^4\big)
\nonumber\\&&        
+2 y^2 (-1+\eta )^2 \big(
        781+7358 \eta +7630 \eta ^2+7358 \eta ^3+781 \eta ^4\big)
\\
R_{52}&=&\eta 
+\eta ^3
+y \eta  \big(
        1+\eta ^2\big)
+y^2 (-1+\eta )^2 \big(
        1+\eta +\eta ^2\big)
\\
R_{53}&=&y (1+\eta ) \Big[
        -y (-1+\eta )^2 (1+\eta )^2
        +y^2 (-1+\eta )^2 \big(
                1+\eta +\eta ^2\big)
\nonumber\\&&                
        -\eta  \big(
                1+\eta +\eta ^2\big)
\Big]\\
R_{54}&=&2 \eta ^3
-y \eta  (1+\eta )^2 \big(
        1-\eta +\eta ^2\big)
+y^3 (-1+\eta )^2 (1+\eta )^2 \big(
        1-\eta +\eta ^2\big)
\nonumber\\&&        
-y^2 (-1+\eta )^2 \big(
        1+2 \eta +2 \eta ^3+\eta ^4\big).
\end{eqnarray}

\section{Numerical Results}
\label{sec:6}

\begin{figure}[ht]
\begin{center}
  \includegraphics[width=.7\textwidth]{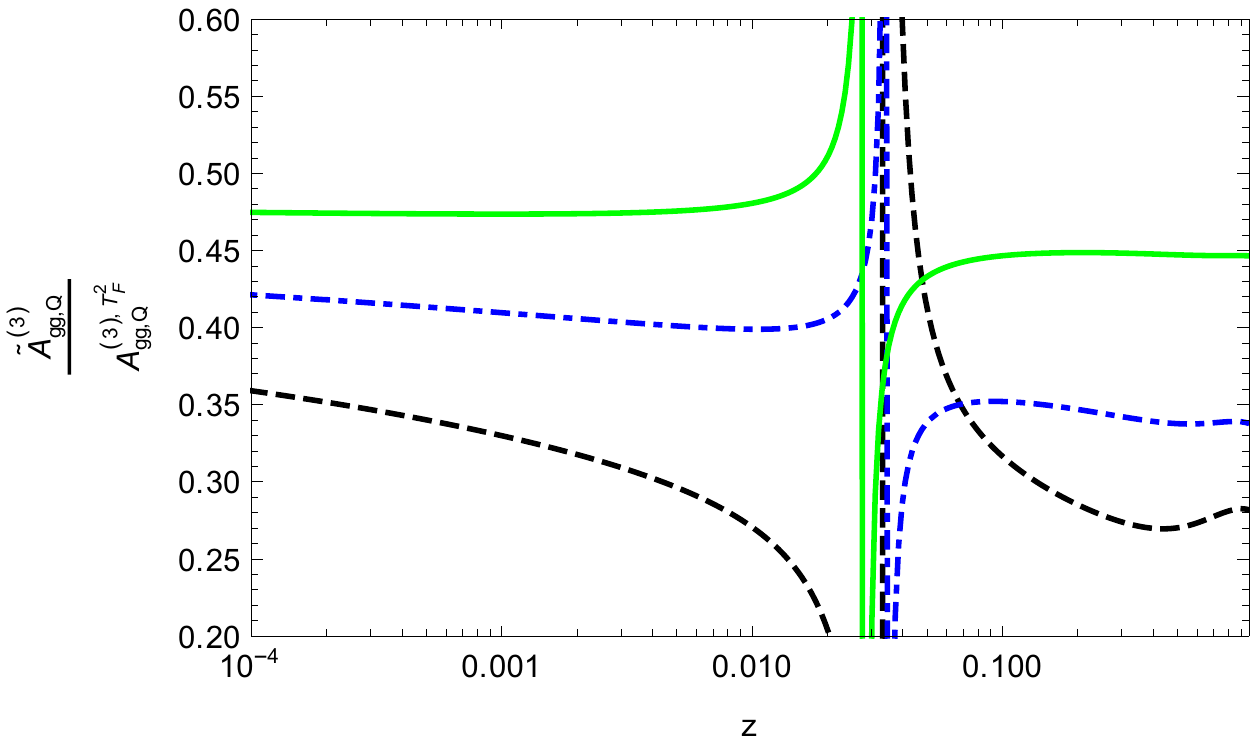}
\caption{\sf%
The ratio of the two mass contribution to the total
contribution at $O(T_F^2)$ for the
polarized massive OME $A_{gg,Q}^{(3)}$ as a function of the momentum
fraction $z$ and the virtuality $\mu^2$.
Dashed line: $\mu^2 = 50~\GeV^2$;
Dash--doted line: $\mu^2 = 100~\GeV^2$;
Full line: $\mu^2 = 1000~\GeV^2$.
For the values of $m_c$ and $m_b$ we refer to the on--shell heavy 
quark masses
$m_c = 1.59~\GeV$ and $m_b = 4.78~\GeV$ \cite{Alekhin:2012vu,PDG}.}
\label{fig:RAT}
\end{center}
\end{figure}

\vspace*{1mm}
\noindent
In Figure~\ref{fig:RAT} we illustrate the size of the two--mass contribution in relation to the total contribution
of $O(T_F^2)$. The polarized single mass corrections to $A_{gg,Q}^{(3)}$ were calculated in \cite{Aggpol3}. Also here
the $N$ space representation has evanescent poles at $N = 1/2, 3/2$, which can be shown to vanish by performing an
analytic expansion.

The corrections are of the size of 20 to 60\%, except of the region $z \sim 0.02 - 0.03$, where the
$O(T_F^2)$ terms 
vanish. In wide ranges the correction behaves as constant for fixed virtualities $\mu^2$. The relative 
size of the 
correction is of similar size as in the unpolarized case \cite{Ablinger:2018brx}.
The two mass corrections are as important as the $O(T_F^2)$ terms and have to be considered
in precision analyses at three--loop order.
\section{Conclusions}
\label{sec:7}
\vspace*{1mm}
\noindent
We have calculated the polarized massive OME $A_{gg,Q}^{(2)}$ in analytic form both in Mellin $N$ space and 
$z$ space in the Larin scheme. In the latter case we made use of a single integral representation, in order 
to shorten the notation, which
would lead to $G$-functions of larger depth for which no numerical representation is available yet. 
The mathematical quantities allowing 
the representation in $N$ space
range up to generalized finite binomial and inverse binomial sums depending on the real parameter $\eta$. In $z$ space
one obtains iterated integrals, based on square--root valued letters, containing the real parameter $\eta$.
The representations derived allow a fast numerical calculation given the analytic representations of the iterative
integrals. The evanescent poles at $N = 1/2$ and $3/2$  present in Mellin $N$ space cancel and the rightmost 
singularity is obtained at $N=0$ as expected.

In the present calculation, the method of direct integration turned out to be the most efficient. Therefore, we did not
perform an integration-by-parts reduction in this two--scale problem, but applied a minimal 
Mellin--Barnes 
representation, followed by the solution of the respective multi--sums using the algorithms encoded in the packages {\tt 
Sigma, EvaluateMultiSums} and {\tt SumProduction}. The obtained expressions have been simplified using the methods
encoded in the package {\tt HarmonicSums}. Checks have been performed using the package {\tt Q2Exp}.
In course of the present calculation we have obtained also a series of new iterative integrals, which could be 
represented in terms of simpler functions. They can be of use in other two--mass calculations.
  
Comparing to the complete $O(T_F^2)$ corrections to $A_{gg,Q}^{(3)}$, we showed that the two--mass contributions form an 
important part and it is necessary to consider them in quantitative analyses. In particular, they contribute to the 
variable flavor number scheme, as well as in other places, at three--loop order. The OME calculated in the present paper
can be used both as a building block for structure functions as well as for the transition matrix in the variable flavor number 
scheme, provided that the Larin scheme is used. There is the possibility also to define the parton distribution functions
and the massless Wilson coefficients in this scheme.

With this result only the two--mass contributions for both the unpolarized and polarized massive 
OME $A_{Qg}^{(3)}$ at three--loop order remain to be calculated, which is work in progress.
\appendix
\section{Representation of certain iterated integrals}
\label{sec:A}

\vspace*{1mm}
\noindent
In the following we present a series of integrals and relations which appeared in intermediate steps of the calculation
and which may be of further use in similar applications, extending the results given in 
\cite{Ablinger:2018brx} before. We obtained the following iterative integrals
\begin{eqnarray}
G_{44} &=& G\left(
        \left\{\frac{\sqrt{1-x}}{x}\right\},z\right) 
=
-\HA_{-1}\big(u_1\big)
-\HA_1\big(u_1\big)
+2 \ln(2)
-2
\nonumber \\ &&
+u_1 \Big[
\HA_0(z) + \HA_1\big(u_1\big) - \HA_{-1}\big(u_1\big) +2
\Big], \\
G_{45} &=& G\left(\left\{\frac{\sqrt{1-x}}{\sqrt{x} (\eta -x \eta +x)}\right\},z\right)
\nonumber \\ &=&
\frac{\pi}{\sqrt{\eta} + \eta} 
+ 2 \frac{{\rm arcsin}(\sqrt{1-z})}{1-\eta} 
- 2 \frac{{\rm arctan}\left(
\frac{\sqrt{\eta} \sqrt{1-z}}{\sqrt{z}}\right)}
{\sqrt{\eta}(1-\eta)}
\\
G_{46} &=& G\left(\left\{\frac{\sqrt{1-x}}{\sqrt{x} (x \eta -x+1)}\right\},z\right)
\nonumber \\ &=&
\frac{1}{\eta -1} \Biggl[\pi  \big(
        \sqrt{\eta }-1\big)
-2 \sqrt{\eta } \, {\rm arctan}\left(\frac{u_1}{\sqrt{\eta z}}\right)
+2 {\rm arctan}\left(\frac{u_1}{\sqrt{z}}\right)
\Biggr], \\
G_{47} &=& G\left(\left\{\frac{\sqrt{x}}{x \eta -x+1}\right\},z\right)
\nonumber \\ &=&
\frac{1}{v_1^3} \Big[
2 \HA_{-1}\big(v_1 \sqrt{z}\big)
+\HA_1(z-z \eta)
\Big]
-\frac{2 \sqrt{z}}{v_1^2}, \\
G_{48} &=& G\left(\left\{\sqrt{1-x} \sqrt{x \eta -x+1}\right\},z\right)
=
\frac{2-\eta}{4 (1-\eta)}
-\frac{u_1^3 u_2}{4}
\nonumber \\ &&
-\frac{u_1 u_2^3}{4 (1-\eta)}
+\frac{\eta^2}{4 v_1^3} \Biggl[
 {\rm arcsinh}\left(\frac{u_1 v_1}{\sqrt{\eta}}\right)
-{\rm arcsinh}\left(\frac{v_1}{\sqrt{\eta}}\right)
\Biggr], \\
G_{49} &=& G\left(\left\{\frac{\sqrt{x \eta -x+1}}{x}\right\},z\right)
=
-\HA_1\big(u_2\big)
-\HA_0(z-z \eta)
-\HA_{-1}\big(u_2\big)
\nonumber \\ &&
+\HA_0(z)
+2 \ln(2)
-2
+u_2 \Big[
         \HA_0(z-z \eta)
        +2
        +\HA_1\big(u_2\big)
        -\HA_{-1}\big(u_2\big)
\Big], \\ 
G_{50} &=& G\left(\left\{\frac{\sqrt{1-x}}{x},\frac{1}{x}\right\},z\right)
=
u_1 \biggl[
         \frac{1}{2} H^2_0(z)
        -4
\biggr]
+2 \ln^2(2)
\nonumber \\ &&
-4 \ln(2)
+4
+\big(1+u_1\big)
\biggl[2 \HA_{-1}\big(u_1\big)
        -\frac{1}{2} \HA_{-1}^2\big(u_1\big)
        +\HA_{-1,1}\big(u_1\big)
\biggr]
\nonumber \\ &&
+\big(1-u_1\big)
\biggl[2 \HA_1\big(u_1\big)
        -\HA_{1,-1}\big(u_1\big)
        +\frac{1}{2} \HA_1^2\big(u_1\big)
\biggr]
-\zeta_2, \\
G_{51} &=& G\left(\left\{\frac{\sqrt{x}}{\eta -x \eta +x},\frac{1}{x}\right\},z\right)
=
       \frac{2 \sqrt{z}}{v_1^2} \big(
                \HA_0(z)
                -2
        \big)
\nonumber \\ &&
+\frac{\sqrt{\eta }}{v_1^3} \Biggl[
        -2 \HA_0(z) \HA_{\{4,0\}}\left(v_1 \sqrt{\frac{z}{\eta}}\right)
        +4 \HA_{\{0,0\},\{4,0\}}\left(v_1 \sqrt{\frac{z}{\eta}}\right)
\Biggr], \\
G_{52} &=& G\left(\left\{\frac{\sqrt{x \eta -x+1}}{x},\frac{1}{x}\right\},z\right)
=
u_2 \big(2 \HA_0(z)-4\big)
+\frac{1}{2} H^2_0(z)
\nonumber \\ &&
+\big(1+u_2\big)
\biggl[-\HA_{-1}\big(u_2\big) \HA_0(z)
       +\HA_{-1,-1}\big(u_2\big)
       -\HA_{1,-1}\big(u_2\big)
\biggr]
\nonumber \\ &&
-2 \HA_0(z)
+\big(1-u_2\big)
\biggl[-\HA_0(z) \HA_1\big(u_2\big)
       -\HA_0(z) \HA_0(z-z \eta)
\nonumber \\ &&
       +\frac{1}{2} \HA_0^2(z-z \eta)
       +\HA_{-1,1}\big(u_2\big)
       -\frac{1}{2} \HA_1^2\big(u_2\big)
\biggr]
+4 \HA_{-1}\big(u_2\big)
-\zeta_2
\nonumber \\ &&
+2 \ln^2(2)
+2 \ln(2) \biggl[\HA_1\big(u_2\big) - \HA_{-1}\big(u_2\big)+\HA_0(z)\biggr]
-4 \ln(2)
+4, \\
G_{53} &=& G\left(\left\{\frac{\sqrt{1-x}}{x},\frac{\sqrt{1-x}}{x},\frac{1}{x}\right\},z\right)
=
z \HA_0(z)
-\frac{z}{2} H^2_0(z)
+\frac{1}{6} H^3_0(z)
\nonumber \\ &&
+2 u_1 \biggl[
        -\frac{1}{2} \HA_{-1}^2\big(u_1\big)
        +\HA_{-1,1}\big(u_1\big)
        -\HA_{1,-1}\big(u_1\big)
       +\frac{1}{2} \HA_1^2\big(u_1\big)
\biggr]
\nonumber \\ &&
-(z+2) \HA_1\big(u_1\big) \HA_{-1}\big(u_1\big)
+\frac{z-2}{2} \Big[\HA_{-1}^2\big(u_1\big)+\HA_1^2\big(u_1\big)\Big]
\nonumber \\ &&
+2 \HA_{-1,1,-1}\big(u_1\big)
-2 \HA_{-1,1,1}\big(u_1\big)
+2 \HA_{1,-1,-1}\big(u_1\big)
-2 \HA_{1,-1,1}\big(u_1\big)
\nonumber \\ &&
+\big(\zeta_2
        +4 \ln(2)
        -2 \ln^2(2)
        +4 u_1
\big)
\Big[
         \HA_{-1}\big(u_1\big)
        +\HA_1\big(u_1\big)
\Big]
-\frac{3}{2} \zeta_3
\nonumber \\ &&
-(3 z+4) \HA_{-1}\big(u_1\big)
+(3 z-4) \HA_1\big(u_1\big)
+\frac{4}{3} \ln^3(2)
\nonumber \\ &&
+\big(
        8
        -8 \ln(2)
        +4 \ln^2(2)
        -2 \zeta_2
\big) u_1
-4 \ln^2(2)
+8 \ln(2)
+6 z
-8, \\
G_{54} &=& G\left(\left\{\frac{1}{x \eta -x+1},\frac{1}{x},\frac{1}{x}\right\},z\right)
=
\frac{1}{1-\eta} \biggl[
\frac{1}{2} H^2_0(z) \HA_1(z-z \eta)
\nonumber \\ &&
-\HA_0(z) \HA_{0,1}(z-z \eta)
+\HA_{0,0,1}(z-z \eta)
\biggr], \\ 
G_{55} &=& G\left(\left\{\frac{\sqrt{x \eta -x+1}}{x},\frac{1}{x},\frac{1}{x}\right\},z\right)
=
-8
+
\frac{4 \ln^3(2)}{3}
\nonumber \\ &&
+\ln(2) \Big[
        8
        -4 \HA_1(\eta )
        -2 \zeta_2
        +\HA_1(\eta )^2
\Big]
\nonumber \\ &&
+u_2 \bigg\{
        8
        +\frac{1}{2} \Big[
                -\HA_1(u_2)
                +\HA_{-1}(u_2)
        \Big]^2 \Big[
                2-\HA_1(\eta )\Big]
\nonumber \\ &&
        -\frac{1}{2} \Big[
                -\HA_1(u_2)
                +\HA_{-1}(u_2)
        \Big]
\Big[-4+\HA_1(\eta )\Big] \HA_1(\eta )
        +\frac{1}{6} \HA_0(z)^3
        -4 \HA_1(\eta )
        -\frac{1}{6} \HA_1(\eta )^3
\nonumber \\ &&
        -\frac{1}{2} \HA_{-1}(u_2) \HA_1(u_2)^2
        +\frac{1}{6} \HA_1(u_2)^3
        -4 \HA_{-1}(u_2)
        -\frac{1}{6} \HA_{-1}(u_2)^3
        +\HA_1(\eta )^2
\nonumber \\ &&
        +\frac{1}{2} \HA_1(u_2) \Big[
                8+\HA_{-1}(u_2)^2\Big]
\bigg\}
+\bigg\{
        4
        +\frac{1}{2} \HA_1(u_2)^2
        +2 \HA_{-1}(u_2)
        -\frac{1}{2} \HA_{-1}(u_2)^2
        -\zeta_2
\nonumber \\ &&
        +\HA_1(u_2) \Big[
                2-\HA_{-1}(u_2)\Big]
\bigg\} \HA_1(\eta )
-\HA_1(\eta )^2
-\frac{1}{2} \Big[
        \HA_1(u_2)
        +\HA_{-1}(u_2)
\Big] \HA_1(\eta )^2
\nonumber \\ &&
+\frac{1}{6} \HA_1(\eta )^3
-\frac{1}{6} \HA_1(u_2)^3
-4 \HA_{-1}(u_2)
-\frac{1}{6} \HA_{-1}(u_2)^3
\nonumber \\ &&
+\Big[
        -4
        +2 \HA_1(\eta )
        +2 \HA_{-1}(u_2)
\Big] \HA_{-1,1}(u_2)
-2 \HA_{-1,1,1}(u_2)
-2 \HA_{-1,-1,1}(u_2)
\nonumber \\ &&
+2 \zeta_2
+2 \zeta_3
+2 \ln^2(2) \Big[
        -2+\HA_1(\eta )\Big]
+\frac{1}{2} \HA_1(u_2)^2 \Big[
        -2+\HA_{-1}(u_2)\Big]
+\HA_{-1}(u_2)^2
\nonumber \\ &&
+\frac{1}{2} \HA_1(u_2) \Big[
        -8+4 \HA_{-1}(u_2)-\HA_{-1}(u_2)^2\Big]
, \\
G_{56} &=& G\left(\left\{\frac{\sqrt{x \eta -x+1}}{x},\frac{1}{x \eta -x+1},\frac{1}{x}\right\},z\right) \nonumber \\
&=&
\frac{u_2}{1-\eta} \biggl\{
         4 \HA_{-1,0}(u_2)
        -4 \HA_{1,0}(u_2)
        -4 \HA_{-1,-1,0}(u_2)
        +4 \HA_{-1,1,0}(u_2)
\nonumber \\ &&
        +\Big[
                4
                -4 \HA_0(u_2)
                +2 \HA_{-1,0}(u_2)
                +\HA_{0,1}(z-z \eta)
                -2 \HA_{1,0}(u_2)
                -\zeta_2
        \Big] \HA_0(z)
\nonumber \\ &&
        -4 \HA_{1,1,0}(u_2)
        -2 \HA_{0,0,1}(z-z \eta)
        +4 \HA_{1,-1,0}(u_2)
        +2 \zeta_2 \HA_{-1}(u_2)
\nonumber \\ &&
        -2 \zeta_2 \HA_1(u_2)
        -2 \zeta_2
        +2 \zeta_3
        -8
\biggr\}
+\frac{1}{1-\eta} \biggl\{
-4 \HA_{-1,-1,0}(u_2)
\nonumber \\ &&
+\Big[
         2 \HA_{-1,0}(u_2)
        +2 \HA_{1,0}(u_2)
        +3 \zeta_2
        -4
\Big] \HA_0(z)
+8
-\frac{7}{2} \zeta_3
\nonumber \\ &&
+4 \HA_{1,1,0}(u_2)
-2 \big(
         \zeta_2
        -4
\big) \HA_{-1}(u_2)
+4 \HA_1(u_2) \zeta_2
-8 \ln(2)
\biggr\}.
\end{eqnarray}
where
\begin{eqnarray}
u_1 &=& \sqrt{1-z}, \\
u_2 &=& \sqrt{1-z (1-\eta)}, \\
v_1 &=& \sqrt{1-\eta}.
\end{eqnarray}




where $C$ is  Catalan's constant and
\begin{eqnarray}
v_2 &=& 1-2 \sqrt{(1-\eta) \eta}, \\
v_3 &=& \frac{v_2}{1-2 \eta}
\end{eqnarray}
and
\begin{eqnarray}
G\left(\left\{\frac{\sqrt{1-x}}{x},\frac{\sqrt{x}}{1-x}\right\},\eta\right) &=& 
\pi^2 - 2 \sqrt{(1 - \eta) \eta} - 2 \arcsin(\sqrt{\eta}) 
\nonumber\\ &&
+
 4 \sqrt{1 - \eta} \arctanh(\sqrt{\eta}) 
+
 8 \arctan\left[-1 + \frac{\sqrt{1 - \eta}}{\sqrt{\eta}}\right] 
\nonumber\\ &&
-
 8 \arctanh\left[\frac{\sqrt{1 - \sqrt{\eta}}}{\sqrt{1 + \sqrt{\eta}}}\right] 
 \arctanh(\sqrt{\eta}) 
\nonumber\\ &&
+
 4 \Li_2\left( -\frac{\sqrt{1 - \sqrt{\eta}}}{\sqrt{1 + \sqrt{\eta}}} \right)
 -
 4 \Li_2\left( \frac{\sqrt{1 - \sqrt{\eta}}}{\sqrt{1 + \sqrt{\eta}}} \right).
\end{eqnarray}
The following constant is calculated numerically
\begin{eqnarray}
G\left(\left\{\frac{\sqrt{2-x}}{x},
\frac{\sqrt{1-x}}{x},
\frac{\sqrt{1-x}}{x}\right\},1\right) &=& 0.413734026910741614953~.
\end{eqnarray}

\section{Relations between certain iterated integrals}
\label{sec:B}

\begin{eqnarray}
G_{59}&=& G \left(\left\{
        \frac{1}{1
        -\tau 
        +\eta  \tau 
        },\frac{1}{\tau },\frac{1}{\tau },\frac{1}{\tau }\right\};x\right)
\nonumber\\
&=& -\frac{1}{1-\eta} \Biggl[
\ln^3(x) \ln(1-(1-\eta)x) + 3 \ln^2(x) \Li_2(x(1-\eta))-6 \ln(x) \Li_3(x(1-\eta)) 
\nonumber\\ && 
+ 6 \Li_4(x(1-\eta))
\Biggr]
\\
G_{60}&=& G \left(\left\{
        \frac{1}{1
        -\tau 
        +\eta  \tau 
        },\frac{1}{\tau },\frac{1}{1-\tau },\frac{1}{1-\tau 
}\right\};1-x\right) = -\frac{1}{1-\eta} \HA_{1/(1-\eta),0,1,1}(1-x)
\\
G_{61}&=& G \left(\left\{
        \frac{1}{1
        -\tau 
        +\eta  \tau 
        },\frac{1}{\tau },\frac{1}{\tau },\frac{1}{1-\tau }\right\};1-x\right) =
-\frac{1}{1-\eta} \HA_{1/(1-\eta),0,0,1}(1-x)
\\
G_{62}&=& G \left(\left\{
        \frac{1}{1
        +\eta 
        -2 \tau 
        +2 \eta  \tau 
        +\tau ^2
        +\eta  \tau ^2
        },\frac{1}{1+\tau }\right\};y\right)
\nonumber\\
&=&-g_{5}(x)
- G \left(\left\{
        \frac{1}{1
        +\eta 
        -2 \tau 
        +2 \eta  \tau 
        +\tau ^2
        +\eta  \tau ^2
        },\frac{1}{1-\tau }\right\};y\right)
\nonumber\\
\\
G_{63}&=& G \left(\left\{
        \frac{1}{1
        +\eta 
        +2 \tau 
        -2 \eta  \tau 
        +\tau ^2
        +\eta  \tau ^2
        },\frac{1}{1+\tau }\right\};y\right)
\nonumber\\
&=&-g_{5}(1-x)
- G \left(\left\{
        \frac{1}{1
        +\eta 
        +2 \tau 
        -2 \eta  \tau 
        +\tau ^2
        +\eta  \tau ^2
        },\frac{1}{1-\tau }\right\};y\right)
\\
G_{64}&=& G \left(\left\{
        \frac{\tau }{1
        +\eta 
        -2 \tau 
        +2 \eta  \tau 
        +\tau ^2
        +\eta  \tau ^2
        },\frac{1}{1+\tau }\right\};y\right)
\nonumber\\
&=&g_{6}(x)
- G \left(\left\{
        \frac{\tau }{1
        +\eta 
        -2 \tau 
        +2 \eta  \tau 
        +\tau ^2
        +\eta  \tau ^2
        },\frac{1}{1-\tau }\right\};y\right)
\\
G_{65}&=& G \left(\left\{
        \frac{\tau }{1
        +\eta 
        +2 \tau 
        -2 \eta  \tau 
        +\tau ^2
        +\eta  \tau ^2
        },\frac{1}{1+\tau }\right\};y\right)
\nonumber\\
&=&-g_{6}(1-x)
- G \left(\left\{
        \frac{\tau }{1
        +\eta 
        +2 \tau 
        -2 \eta  \tau 
        +\tau ^2
        +\eta  \tau ^2
        },\frac{1}{1-\tau }\right\};y\right)
\\
G_{66}&=& G \left(\left\{
        \frac{1}{1
        +\eta 
        -2 \tau 
        +2 \eta  \tau 
        +\tau ^2
        +\eta  \tau ^2
        },\frac{\tau }{1+\tau ^2}\right\};y\right)
\nonumber\\
&=&g_{7}(x)
- G \left(\left\{
        \frac{1}{1
        +\eta 
        -2 \tau 
        +2 \eta  \tau 
        +\tau ^2
        +\eta  \tau ^2
        },\frac{1}{1-\tau }\right\};y\right)
\\
G_{67}&=& G \left(\left\{
        \frac{1}{1
        +\eta 
        +2 \tau 
        -2 \eta  \tau 
        +\tau ^2
        +\eta  \tau ^2
        },\frac{\tau }{1+\tau ^2}\right\};y\right)
\nonumber\\
&=&g_{8}(x)
- G \left(\left\{
        \frac{1}{1
        +\eta 
        +2 \tau 
        -2 \eta  \tau 
        +\tau ^2
        +\eta  \tau ^2
        },\frac{1}{1-\tau }\right\};y\right)
\\
G_{68}&=& G \left(\left\{
        \frac{1}{1
        +\eta 
        +2 \tau 
        -2 \eta  \tau 
        +\tau ^2
        +\eta  \tau ^2
        },\frac{1}{1+\tau ^2}\right\};y\right)
= g_{9}(x)\\
G_{69}&=& G \left(\left\{
        \frac{1}{1
        +\eta 
        -2 \tau 
        +2 \eta  \tau 
        +\tau ^2
        +\eta  \tau ^2
        },\frac{1}{1+\tau ^2}\right\};y\right)
= g_{9}(1-x)\\
G_{70}&=& G \left(\left\{
        \frac{\tau }{1
        +\eta 
        -2 \tau 
        +2 \eta  \tau 
        +\tau ^2
        +\eta  \tau ^2
        }\right\};y\right)
= g_{10}(x)\\
G_{71}&=& G \left(\left\{
        \frac{\tau }{1
        +\eta 
        +2 \tau 
        -2 \eta  \tau 
        +\tau ^2
        +\eta  \tau ^2
        }\right\};y\right)
= g_{10}(1-x)\\
G_{72}&=& G \left(\left\{
        \frac{1}{1
        +\eta 
        +2 \tau 
        -2 \eta  \tau 
        +\tau ^2
        +\eta  \tau ^2
        }\right\};y\right)
= g_{11}(x)\\
G_{73}&=& G \left(\left\{
        \frac{1}{1
        +\eta 
        -2 \tau 
        +2 \eta  \tau 
        +\tau ^2
        +\eta  \tau ^2
        }\right\};y\right)
= -g_{11}(1-x)\\
G_{74}&=& G \left(\left\{
        \frac{1}{1
        -\tau 
        +\eta  \tau 
        },\frac{1}{\tau },\frac{1}{1-\tau }\right\};x\right) = \frac{1}{1-\eta} \HA_{1/(1-\eta),0,1}(x)
\nonumber\\ &=&
g_{12}(x)
- G \left(\left\{
        \frac{1}{1
        -\tau 
        +\eta  \tau 
        },\frac{1}{1-\tau },\frac{1}{\tau }\right\};x\right)
\\
G_{75}&=& G \left(\left\{
        \frac{1}{1
        -\tau 
        +\eta  \tau 
        },\frac{1}{\tau },\frac{1}{1-\tau }\right\};1-x\right) = \frac{1}{1-\eta} \HA_{1/(1-\eta),0,1}(1-x)
\nonumber\\
&=&g_{12}(1-x)
- G \left(\left\{
        \frac{1}{1
        -\tau 
        +\eta  \tau 
        },\frac{1}{1-\tau },\frac{1}{\tau }\right\};1-x\right) 
\\
G_{76}&=& G \left(\left\{
        \frac{1}{1
        -\tau 
        +\eta  \tau 
        },\frac{1}{1-\tau },\frac{1}{\tau }\right\};x\right) = \frac{1}{1-\eta} \HA_{1/(1-\eta),1,0}(x)
= g_{13}(x).
\end{eqnarray}
Some of the integrals are generalized harmonic polylogarithms based on three different letters, which are
known not to reduce to Nielsen integrals \cite{NIELSEN} in general. Here one of the letters contains a linear 
function of the real parameter $\eta$.

The functions $G_{62}$ to $G_{67}$ contain the $\eta$-dependent letters
\begin{eqnarray}
\label{eq:qf1}
&& \frac{1}{(1-t)^2 + (1+t)^2 \eta},
\\
&& \frac{t}{(1-t)^2 + (1+t)^2 \eta},
\end{eqnarray}
and their replacement with $\eta \rightarrow 1/\eta$. These letters can be partial fractioned as
\begin{eqnarray}
\frac{1}{(t-a)(t-b)} &=& \frac{1}{a-b} \left[\frac{1}{t-a} - \frac{1}{t-b} \right]
\\
\frac{t}{(t-a)(t-b)} &=& \frac{1}{a-b} \left[\frac{a}{t-a} - \frac{b}{t-b}\right].
\\
\end{eqnarray}
Then these functions can be obtained as linear combinations of the following four integrals
\begin{eqnarray}
I_1 &=& \int_0^y dt \frac{\ln(1+t)}{t-a} = \ln\left[\frac{a - y}{a+1}\right] \ln(1+y) 
- \Li_2\left[\frac{1}{1+a}\right]
+ \Li_2\left[\frac{1+y}{1+a}\right]
\\
I_2 &=& -\int_0^y \frac{\ln(1-t)}{t-a} = - \ln(1 - y) \ln\left[\frac{a - y}{a-1}\right] 
+ \Li_2\left[\frac{1}{1-a}\right]
- \Li_2\left[\frac{y-1}{a-1}\right]
\\
I_3 &=& \int_0^y \frac{\arctan(t)}{t - a} = -\frac{i}{2}
\Biggl\{
2 i \arctan{y} \ln(a-y) + \ln(i + a) \ln(1- iy) 
- \ln(a-i) \ln(1+iy) 
\nonumber\\ &&
- \Li_2\left[-\frac{i}{a-i}\right]
+ \Li_2\left[\frac{i}{i+a}\right]
+ \Li_2\left[\frac{y-i}{a-i}\right]
- \Li_2\left[\frac{y+i}{a+i}\right]
\Biggr\}
\\
I_4 &=& \frac{1}{2} \int_0^y \frac{\ln(1+t^2)}{t-a} = \frac{1}{2} \Biggl\{
-\left[\ln(a-i) + \ln(a+i)\right] \left[\ln(-a) - \ln(y-a)\right] 
\nonumber\\ &&
+ \Li_2\left[\frac{a}{a-i}\right] 
+ \Li_2\left[\frac{a}{a+i}\right] 
- \Li_2\left[\frac{a - y}{a-i}\right] 
- \Li_2\left[\frac{a - y}{a+i}\right]
\Biggr\}.
\end{eqnarray}
The roots of the quadratic form (\ref{eq:qf1}) 
\begin{eqnarray}
(1-t)^2+(1+t^2) \eta =  c (t-a)(t-b)
\end{eqnarray}
are:
\begin{eqnarray}
a &=& \frac{1 - 2i \sqrt{\eta} - \eta}{1+\eta},
\\
b &=& \frac{1 + 2i \sqrt{\eta} - \eta}{1+\eta},
\end{eqnarray}
with
\begin{eqnarray}
c &=& 1 + \eta.
\end{eqnarray}

We further present representations of a series of functions $g_i$ which are functions of $x$ and $\eta$. 
The symbol $y$, not to be confused with its meaning in the main text, is defined here as
\begin{equation}
y = \frac{1-2 \sqrt{1-x} \sqrt{x}}{1-2 x},
\end{equation}
and the formulas are valid for
\begin{eqnarray}
0 & <\eta &<1, \nonumber\\
0 & <x    &<1, 
\end{eqnarray}


{\bf Acknowledgments}

\noindent
This project has received funding from the European Union's Horizon 2020 research and innovation programme under the 
Marie Sk\l{}odowska-Curie grant agreement No. 764850, SAGEX, and COST action CA16201: Unraveling new physics at the LHC 
through the precision frontier and from the Austrian Science Fund (FWF) grant SFB F50 (F5009-N15).
The diagrams have been drawn using {\tt Axodraw} \cite{Vermaseren:1994je}.


\begin{thebibliography}{99}
%
\bibitem{Buza:1995ie}
  M.~Buza, Y.~Matiounine, J.~Smith, R.~Migneron and W.L.~van Neerven,
  Nucl.\ Phys.\ B {\bf 472} (1996) 611--658
  [hep-ph/9601302].
%
\bibitem{Ablinger:2014vwa}
  J.~Ablinger {\it et al.},
  Nucl.\ Phys.\ B {\bf 886} (2014) 733--823
  [arXiv:1406.4654 [hep-ph]].
%
\bibitem{Buza:1996xr}
  M.~Buza, Y.~Matiounine, J.~Smith and W.~L.~van Neerven,
  Nucl.\ Phys.\ B {\bf 485} (1997) 420--456
  [hep-ph/9608342].
%
\bibitem{Buza:1996wv}
  M.~Buza, Y.~Matiounine, J.~Smith and W.~L.~van Neerven,
  Eur.\ Phys.\ J.\ C {\bf 1} (1998) 301--320
  [hep-ph/9612398].
%
\bibitem{Bierenbaum:2007qe}
  I.~Bierenbaum, J.~Bl\"umlein and S.~Klein,
  Nucl.\ Phys.\ B {\bf 780} (2007) 40--75
  [hep-ph/0703285].
%
\bibitem{Bierenbaum:2007pn}
  I.~Bierenbaum, J.~Bl\"umlein and S.~Klein,
  {\it Two-loop massive operator matrix elements for polarized and unpolarized deep-inelastic scattering},
  arXiv:0706.2738 [hep-ph].
%
\bibitem{Bierenbaum:2008yu}
  I.~Bierenbaum, J.~Bl\"umlein, S.~Klein and C.~Schneider,
  Nucl.\ Phys.\ B {\bf 803} (2008) 1--41
  [arXiv:0803.0273 [hep-ph]].
%
\bibitem{Ablinger:2010ty}
  J.~Ablinger, J.~Bl\"umlein, S.~Klein, C.~Schneider and F.~Wi\ss{}brock,
  Nucl.\ Phys.\ B {\bf 844} (2011) 26--54
  [arXiv:1008.3347 [hep-ph]].
%
\bibitem{Ablinger:2014nga}
  J.~Ablinger, A.~Behring, J.~Bl\"umlein, A.~De Freitas, A.~von Manteuffel and C.~Schneider,
  Nucl.\ Phys.\ B {\bf 890} (2014) 48--151
  [arXiv:1409.1135 [hep-ph]].
%
\bibitem{Ablinger:2014uka}
  J.~Ablinger, J.~Bl\"umlein, A.~De Freitas, A.~Hasselhuhn, A.~von Manteuffel, M.~Round and C.~Schneider,
  Nucl.\ Phys.\ B {\bf 885} (2014) 280--317
  [arXiv:1405.4259 [hep-ph]].
%
\bibitem{Behring:2014eya}
  A.~Behring, I.~Bierenbaum, J.~Bl\"umlein, A.~De Freitas, S.~Klein and F.~Wi\ss{}brock,
  Eur.\ Phys.\ J.\ C {\bf 74} (2014) no.9,  3033
  [arXiv:1403.6356 [hep-ph]].
%
\bibitem{Blumlein:2014fqa}
  J.~Bl\"umlein, A.~Hasselhuhn and T.~Pfoh,
  Nucl.\ Phys.\ B {\bf 881} (2014) 1--41
  [arXiv:1401.4352 [hep-ph]].
%
\bibitem{Ablinger:2014lka}
  J.~Ablinger, J.~Bl\"umlein, A.~De Freitas, A.~Hasselhuhn, A.~von Manteuffel, M.~Round, C.~Schneider and F.~Wi\ss{}brock,
  Nucl.\ Phys.\ B {\bf 882} (2014) 263--288
  [arXiv:1402.0359 [hep-ph]].
%
\bibitem{Ablinger:2017ptf}
  J.~Bl\"umlein, J.~Ablinger, A.~Behring, A.~De Freitas, A.~von Manteuffel, C.~Schneider and C.~Schneider,
  PoS (QCDEV2017) 031
  [arXiv:1711.07957 [hep-ph]].
%
\bibitem{Ablinger:2016kgz}
  J.~Ablinger {\it et al.},
  PoS (QCDEV2016)  052
  [arXiv:1611.01104 [hep-ph]].
%
\bibitem{Behring:2016hpa}
  A.~Behring, J.~Bl\"umlein, G.~Falcioni, A.~De Freitas, A.~von Manteuffel and C.~Schneider,
  Phys.\ Rev.\ D {\bf 94} (2016) no.11,  114006
  [arXiv:1609.06255 [hep-ph]].
%
\bibitem{Blumlein:2016xcy}
  J.~Bl\"umlein, G.~Falcioni and A.~De Freitas,
  Nucl.\ Phys.\ B {\bf 910} (2016) 568--617
  [arXiv:1605.05541 [hep-ph]].
%
\bibitem{Behring:2015zaa}
  A.~Behring, J.~Bl\"umlein, A.~De Freitas, A.~von Manteuffel and C.~Schneider,
  Nucl.\ Phys.\ B {\bf 897} (2015) 612--644
  [arXiv:1504.08217 [hep-ph]].
%
\bibitem{Ablinger:2019etw}
  J.~Ablinger, A.~Behring, J.~Bl\"umlein, A.~De Freitas, A.~von Manteuffel, C.~Schneider and K.~Sch\"onwald,
  Nucl.\ Phys.\ B {\bf 953} (2020) 114945
  [arXiv:1912.02536 [hep-ph]].
%
\bibitem{Ablinger:2019gpu}
  J.~Ablinger, J.~Bl\"umlein, A.~De Freitas, M.~Saragnese, C.~Schneider and K.~Sch\"onwald,
  Nucl.\ Phys.\ B {\bf 952} (2020) 114916
  [arXiv:1911.11630 [hep-ph]].
%
\bibitem{Blumlein:2019zux}
  J.~Bl\"umlein, C.~Raab and K.~Sch\"onwald,
  Nucl.\ Phys.\ B {\bf 948} (2019) 114736
  [arXiv:1904.08911 [hep-ph]].
%
\bibitem{Blumlein:2019qze}
  J.~Bl\"umlein, A.~De Freitas, C.G.~Raab and K.~Sch\"onwald,
  Nucl.\ Phys.\ B {\bf 945} (2019) 114659
  [arXiv:1903.06155 [hep-ph]].
%
\bibitem{Blumlein:2018jfm}
  J.~Bl\"umlein, A.~De Freitas, C.~Schneider and K.~Sch\"onwald,
  Phys.\ Lett.\ B {\bf 782} (2018) 362--366
  [arXiv:1804.03129 [hep-ph]].
%
\bibitem{Ablinger:2017xml}
  J.~Ablinger, J.~Bl\"umlein, A.~De Freitas, C.~Schneider and K.~Sch\"onwald,
  Nucl.\ Phys.\ B {\bf 927} (2018) 339--367
  [arXiv:1711.06717 [hep-ph]].
%
\bibitem{Ablinger:2018brx}
  J.~Ablinger, J.~Bl\"umlein, A.~De Freitas, A.~Goedicke, C.~Schneider and K.~Sch\"onwald,
  Nucl.\ Phys.\ B {\bf 932} (2018) 129--230
  [arXiv:1804.02226 [hep-ph]].
%
\bibitem{Ablinger:2017err}
  J.~Ablinger, J.~Bl\"umlein, A.~De Freitas, A.~Hasselhuhn, C.~Schneider and F.~Wi\ss{}brock,
  Nucl.\ Phys.\ B {\bf 921} (2017) 585--688
  [arXiv:1705.07030 [hep-ph]].
%
\bibitem{Ablinger:2012qj}
  J.~Ablinger, J.~Bl\"umlein, A.~Hasselhuhn, S.~Klein, C.~Schneider and F.~Wi\ss{}brock,
  PoS (RADCOR2011) 031
  [arXiv:1202.2700 [hep-ph]].
%
\bibitem{Ablinger:2011pb}
  J.~Ablinger, J.~Bl\"umlein, S.~Klein, C.~Schneider and F.~Wi\ss{}brock,
  {\it 3-Loop Heavy Flavor Corrections to DIS with two Massive Fermion Lines},
  arXiv:1106.5937 [hep-ph].
%
\bibitem{Ablinger:2017tan}
  J.~Ablinger, A.~Behring, J.~Bl\"umlein, A.~De Freitas, A.~von Manteuffel and C.~Schneider,
  Nucl.\ Phys.\ B {\bf 922} (2017) 1--40
  [arXiv:1705.01508 [hep-ph]].
%
\bibitem{Behring:2019tus}
  A.~Behring, J.~Bl\"umlein, A.~De Freitas, A.~Goedicke, S.~Klein, A.~von Manteuffel, C.~Schneider and K.~Sch\"onwald,
  Nucl.\ Phys.\ B {\bf 948} (2019) 114753
  [arXiv:1908.03779 [hep-ph]].
%
\bibitem{Moch:2004pa}
  S.~Moch, J.A.M.~Vermaseren and A.~Vogt,
  Nucl.\ Phys.\ B {\bf 688} (2004) 101--134
  [hep-ph/0403192].
%
\bibitem{Vogt:2004mw}
  A.~Vogt, S.~Moch and J.A.M.~Vermaseren,
  Nucl.\ Phys.\ B {\bf 691} (2004) 129--181
  [hep-ph/0404111].
%
\bibitem{Moch:2014sna}
  S.~Moch, J.A.M.~Vermaseren and A.~Vogt,
  Nucl.\ Phys.\ B {\bf 889} (2014) 351--400
  [arXiv:1409.5131 [hep-ph]].
%
\bibitem{SIG1}
C.~Schneider, {S\'em.~Lothar. Combin.\/} {\bf 56} (2007) 1--36, 
 article B56b.
%
\bibitem{SIG2}
C.~Schneider, {{\sf Computer Algebra in Quantum Field Theory: Integration,
  Summation and Special Functions}\/} Texts and Monographs in Symbolic
  Computation, eds. C.~Schneider and J.~Bl\"umlein  (Springer, Wien, 2013) 325-360
  {[arXiv:1304.4134 [cs.SC]]}.
%
\bibitem{Vermaseren:1998uu}
  J.A.M.~Vermaseren,
  Int.\ J.\ Mod.\ Phys.\ A {\bf 14} (1999) 2037--2076
  [hep-ph/9806280].
%
\bibitem{Blumlein:1998if}
  J.~Bl\"umlein and S.~Kurth,
  Phys.\ Rev.\ D {\bf 60} (1999) 014018
  [hep-ph/9810241].
%
\bibitem{Ablinger:2013cf}
  J.~Ablinger, J.~Bl\"umlein and C.~Schneider,
  J.\ Math.\ Phys.\  {\bf 54} (2013) 082301
  [arXiv:1302.0378 [math-ph]].
%
\bibitem{Ablinger:2011te}
  J.~Ablinger, J.~Bl\"umlein and C.~Schneider,
  J.\ Math.\ Phys.\  {\bf 52} (2011) 102301
  [arXiv:1105.6063 [math-ph]].
%
\bibitem{Ablinger:2014bra}
  J.~Ablinger, J.~Bl\"umlein, C.G.~Raab and C.~Schneider,
  J.\ Math.\ Phys.\  {\bf 55} (2014) 112301
  [arXiv:1407.1822 [hep-th]].
%
\bibitem{Ablinger:2010kw}
  J.~Ablinger,
    {\it A Computer Algebra Toolbox for Harmonic Sums Related to Particle Physics}, Diploma Thesis, J. Kepler 
University Linz, 2009,
  arXiv:1011.1176 [math-ph].
%
\bibitem{Ablinger:2013hcp}
  J.~Ablinger,
  {\it Computer Algebra Algorithms for Special Functions in Particle Physics}, Ph.D. Thesis, J. Kepler University Linz, 2012,
  arXiv:1305.0687 [math-ph].
%
\bibitem{Ablinger:2014rba}
  J.~Ablinger,
  PoS (LL2014) 019
  [arXiv:1407.6180 [cs.SC]].
%
\bibitem{EMSSP}
  J.~Ablinger, J.~Bl\"umlein, S.~Klein and C.~Schneider,
  Nucl.\ Phys.\ Proc.\ Suppl.\  {\bf 205-206} (2010) 110--115
  [arXiv:1006.4797 [math-ph]];\\
  J.~Bl\"umlein, A.~Hasselhuhn and C.~Schneider,
  PoS (RADCOR2011) 032
  [arXiv:1202.4303 [math-ph]];\\
  C.~Schneider,
  J.\ Phys.\ Conf.\ Ser.\  {\bf 523} (2014) 012037
  [arXiv:1310.0160 [cs.SC]].
%
\bibitem{Larin:1993tq}
  S.A.~Larin,
  Phys.\ Lett.\ B {\bf 303} (1993) 113--118
  [hep-ph/9302240].
%
\bibitem{Blumlein:2018cms}
  J.~Bl\"umlein and C.~Schneider,
  Int.\ J.\ Mod.\ Phys.\ A {\bf 33} (2018) no.17,  1830015
  [arXiv:1809.02889 [hep-ph]].
%
\bibitem{Moch:2001zr}
  S.~Moch, P.~Uwer and S.~Weinzierl,
  J.\ Math.\ Phys.\  {\bf 43} (2002) 3363--3386
  [hep-ph/0110083].
%
\bibitem{Remiddi:1999ew}
  E.~Remiddi and J.A.M.~Vermaseren,
  Int.\ J.\ Mod.\ Phys.\ A {\bf 15} (2000) 725--754
  [hep-ph/9905237].
%
\bibitem{Ablinger:2016lzr}
  J.~Ablinger,
  PoS (LL2016) 067.
%
\bibitem{Ablinger:2018cja}
  J.~Ablinger,
  PoS (RADCOR2017) 001.
%
\bibitem{Ablinger:2018pwq}
  J.~Ablinger,
  PoS (LL2018) 063.
%
\bibitem{Blumlein:2017mtk}
  J.~Bl\"umlein, A.~De Freitas, C.~Schneider and K.~Sch\"onwald,
  {\it The Three Loop Two-Mass Contribution to the Gluon Vacuum Polarization},
  arXiv:1710.04500 [hep-ph].
%
\bibitem{QGRAF}
  P.~Nogueira,
  J.\ Comput.\ Phys.\  {\bf 105} (1993) 279--289.
%
\bibitem{FORM}
  J.A.M.~Vermaseren,
  {\it New features of FORM},
  math-ph/0010025;\\
  M.~Tentyukov and J.A.M.~Vermaseren,
  Comput.\ Phys.\ Commun.\  {\bf 181} (2010) 1419--1427
  [hep-ph/0702279].
%
\bibitem{Mertig:1995ny}
  R.~Mertig and W.~L.~van Neerven,
  Z.\ Phys.\ C {\bf 70} (1996) 637--653
  [hep-ph/9506451].
%
\bibitem{Yndurain:1999ui}
  F.~J.~Yndurain,
  {\sf The theory of quark and gluon interactions},
  (Springer, Berlin, 2006).
%
\bibitem{Bierenbaum:2009mv}
  I.~Bierenbaum, J.~Bl\"umlein and S.~Klein,
  Nucl.\ Phys.\ B {\bf 820} (2009) 417--482
  [arXiv:0904.3563 [hep-ph]].
%
\bibitem{Klein:2009ig}
  S.W.G.~Klein,
  {\it Mellin Moments of Heavy Flavor Contributions to $F_2(x,Q^2)$ at NNLO}, Ph.D. Thesis, TU Dortmund (2009)
  arXiv:0910.3101 [hep-ph].
%
\bibitem{Blumlein:2009cf}
J.~Bl\"umlein, D.~Broadhurst and J.A.M.~Vermaseren,
Comput.\ Phys.\ Commun.\  \textbf{181} (2010), 582--625
[arXiv:0907.2557 [math-ph]].
%
\bibitem{MB1a}
E.W.~Barnes, 
Proc. Lond. Math. Soc. (2) {\bf 6} (1908) 141--177.
%
\bibitem{MB1b}
E.W.~Barnes,
Quarterly Journal of Mathematics {\bf 41} (1910) 136--140.
%
\bibitem{MB2}
H.~Mellin,
Math. Ann. {\bf 68}, no. 3 (1910) 305--337.
%
\bibitem{MB3}
E.T.~Whittaker and G.N.~Watson, {\sf A Course of Modern Analysis}, (Cambridge University Press, Cambridge, 1927;
                   reprinted 1996). 
%
\bibitem{MB4}
E.C.~Titchmarsh,
{\sf Introduction to the Theory of Fourier Integrals},
(Calendron Press, Oxford, 1937; 2nd Edition 1948).
%
\bibitem{MB}
  M.~Czakon,
  Comput.\ Phys.\ Commun.\  {\bf 175} (2006) 559--571
  [hep-ph/0511200].
%
\bibitem{MBr}
  A.V.~Smirnov and V.A.~Smirnov,
  Eur.\ Phys.\ J.\ C {\bf 62} (2009) 445--449
  [arXiv:0901.0386 [hep-ph]].
%
\bibitem{Ablinger:2013jta}
  J.~Ablinger and J.~Bl\"umlein,
  {\it Harmonic Sums, Polylogarithms,Special Numbers, and Their Generalizations},
  arXiv:1304.7071 [math-ph].
%
\bibitem{Harlander:1997zb}
  R.~Harlander, T.~Seidensticker and M.~Steinhauser,
  Phys.\ Lett.\ B {\bf 426} (1998) 125--132
  [hep-ph/9712228].
%
\bibitem{Seidensticker:1999bb}
  T.~Seidensticker,
  {\it Automatic application of successive asymptotic expansions of Feynman diagrams},
  hep-ph/9905298.
%
\bibitem{Aggpol3}
J.~Ablinger, A.~Behring, J.~Bl\"umlein, A.~De Freitas,  A.~Goedicke, A.~von Manteuffel,
and C.~Schneider, DESY 20--053.
%
\bibitem{Alekhin:2012vu}
  S.~Alekhin, J.~Bl\"umlein, K.~Daum, K.~Lipka and S.~Moch,
  Phys.\ Lett.\ B {\bf 720} (2013) 172--176
  [arXiv:1212.2355 [hep-ph]].
%
\bibitem{PDG}
K.A.~Olive et al. [Particle Data Group], Chin. Phys. C {\bf 38} (2014) 090001.
%
\bibitem{NIELSEN}
N.~Nielsen,
Nova Acta
Leopold. {\bf XC} (1909) Nr. 3, 125--211;\\
  K.S.~K\"olbig,
  SIAM J.\ Math.\ Anal.\  {\bf 17} (1986) 1232--1258;\\
  A.~Devoto and D.W.~Duke,
  Riv.\ Nuovo Cim.\  {\bf 7N6} (1984) 1--39;\\
L. Lewin, {\sf Dilogarithms and associated functions} (Macdonald, London, 1958);\\
L. Lewin, {\sf Polylogarithms and Associated Functions}, (North Holland, Amsterdam, 1981).
%
\bibitem{Vermaseren:1994je}
  J.A.M.~Vermaseren,
  Comput.\ Phys.\ Commun.\  {\bf 83} (1994) 45--58.
\end{thebibliography}
\end{document}